\documentclass[fleqn,10pt]{wlscirep}
\usepackage{subfigure}
\usepackage[utf8]{inputenc}
\usepackage[T1]{fontenc}
\usepackage{lineno}
\usepackage{mhchem}
\usepackage{cleveref}
\usepackage{multirow}
\crefrangeformat{table}{#3#1#4--#5#2#6}

\title{Interpretable Machine Learning for Quantum-\\Informed Property Predictions in Artificial Sensing Materials}

\def\etal{{\it et al. }}
\newcommand{\ie}[0]{\textit{i.e.}, }

\newcommand{\eg}[0]{\textit{e.g.}, }

\newcommand{\DefineAuthor}[2]{%
  \expandafter\newcommand\csname #1note\endcsname[1]{%
    \textbf{\textcolor{#2}{\textbf{#1:} ##1}}}%
  \expandafter\newcommand\csname #1\endcsname[1]{
    \textbf{\textcolor{#2}{##1}}}
  \expandafter\newcommand\csname #1cancel\endcsname[1]{%
    \textbf{\textcolor{#2}{\sout{##1}}}}%
  \expandafter\newcommand\csname #1change\endcsname[2]{%
    \textbf{\textcolor{#2}{\sout{##1} ##2}}}%
  \newenvironment{#1text}{\color{#2}}{\color{black}}
}
\definecolor{dartmouthgreen}{rgb}{0.05, 0.5, 0.06}
\DefineAuthor{LMS}{red}
\DefineAuthor{Li}{blue}
\DefineAuthor{ma}{orange}

\author[1]{Li Chen}
\author[1,*]{Leonardo Medrano Sandonas}
\author[1]{Shirong Huang}
\author[2]{Alexander Croy}
\author[1,3,4,5,*]{Gianaurelio Cuniberti}
\affil[1]{Institute for Materials Science and Max Bergmann Center for Biomaterials, TUD Dresden University of Technology, 01062 Dresden, Germany}
\affil[2]{Institute of Physical Chemistry, Friedrich Schiller University Jena, 07737 Jena, Germany}
\affil[3]{Dresden Center for Computational Materials Science (DCMS), TUD Dresden University of Technology, 01062 Dresden, Germany}
\affil[4]{Cluster of Excellence CARE, TU Dresden and RWTH Aachen, Germany}
\affil[5]{Cluster of Excellence CeTI, TU Dresden, Germany}
\affil[*]{corresponding author(s): Leonardo Medrano Sandonas (leonardo.medrano@tu-dresden.de), Gianaurelio Cuniberti (gianaurelio.cuniberti@tu-dresden.de)}

\begin{abstract} 
Digital sensing faces challenges in developing sustainable methods to extend the applicability of customized e-noses to complex body odor volatilome (BOV). To address this challenge, we developed MORE-ML, a computational framework that integrates quantum-mechanical (QM) property data of e-nose molecular building blocks with machine learning (ML) methods to predict sensing-relevant properties. Within this framework, we expanded our previous dataset, MORE-Q, to MORE-QX by sampling a larger conformational space of interactions between BOV molecules and mucin-derived receptors. This dataset provides extensive electronic binding features (BFs) computed upon BOV adsorption. Analysis of MORE-QX property space revealed weak correlations between QM properties of building blocks and resulting BFs. Leveraging this observation, we defined electronic descriptors of building blocks as inputs for tree-based ML models to predict BFs. Benchmarking showed CatBoost models outperform alternatives, especially in transferability to unseen compounds. Explainable AI methods further highlighted which QM properties most influence BF predictions. Collectively, MORE-ML combines QM insights with ML to provide mechanistic understanding and rational design principles for molecular receptors in BOV sensing. This approach establishes a foundation for advancing artificial sensing materials capable of analyzing complex odor mixtures, bridging the gap between molecular-level computations and practical e-nose applications.

\end{abstract}

\begin{document}

\flushbottom
\maketitle

\thispagestyle{empty}

\section*{Introduction}
The rapid advancement in artificial intelligence (AI) has significantly accelerated the development of AI-driven technologies, enabling precise recognition of objects, faces, voices, and tactile sensations \cite{zhang2020emotion,wang2021artificial_sense}. 
Despite these advancements, a considerable technological gap persists in effectively interpreting and predicting the chemical environment surrounding humans. 
To bridge this gap, customized electronic noses have emerged, demonstrating notable proficiency in detecting volatile organic compounds (VOCs) \cite{ali2020_enose,huang2021_enose,huang2022_enose,li2024_voc_enose}. 
Specifically, VOCs emitted from the human body (referred to as body odor volatilome (BOV)) act as unique chemical fingerprints and hold great promise for healthcare applications \cite{drabinska_bovs}, \eg serving as biomarkers for Alzheimer's and Parkinson's diseases\cite{trivedi2019_parkinson,tisch2013detection}.
However, there remains a strong and persistent need for rapid and reliable sensing materials capable of detecting biomarkers\cite{karnaushenko2015light_biomakers} \eg BOV molecules within digital olfactory systems, particularly for medical diagnostics.

Inspired by the sensitivity \cite{2001_OR} and the discriminative power of the human olfactory system \cite{2016_sp_smeller}, diverse molecular olfactory receptors have recently been synthesized (\eg mucin-derived receptors\cite{bakhatan2023_rec1,sukhran2023_rec2,Huang2023}).
This progress has driven the development of experimental protocols aimed at controlling receptor affinity toward BOV molecules in gas sensing by incorporating specific functional  groups with varying chemical characteristics on glaco-conjugated\cite{shitrit2025monosaccharide}.
However, obtaining detailed information on BOV–receptor interactions---and thus guidance for receptor optimization---remains both costly and time-consuming when relying on empirical trial-and-error screening.
This indicates that a key limitation of current prototype receptors lies in the lack of mechanistic insight into their sensitivity and selectivity across the vast chemical space of BOV–receptor systems.
This bottleneck underscores the need for sustainable strategies to rationally design high-performance receptor-based biomimetic sensors.
Similar to the transformative impact of molecular electronics a few decades ago\cite{cuniberti2005_molecular_electronics}, quantum-mechanical (QM) methodologies could revolutionize the field of chemical sensing by providing a deeper understanding of the physical and chemical interactions that govern key performance metrics such as recovery time, charge transfer, and Schottky barrier potential\cite{bhati2021gas_sensing_mechanism}.
Furthermore, integrating QM-derived property data with AI techniques has the potential to yield reliable and efficient computational frameworks for guiding the design of materials for sensors with high sensitivity and selectivity---an approach that has recently proven successful in drug discovery studies\cite{Ginex2024,Vargas2025,hinostroza25,Manathunga2022}.

Within this context, we have recently introduced the MORE-Q dataset \cite{more-q}, providing, for the first time, an extensive set of QM property data corresponding to the atomistic building blocks of artificial olfactory molecular sensors: BOV molecules, mucin-derived olfactorial receptors, and BOV-receptor dimer systems. 
MORE-Q also contains electronic structure data describing the intermolecular interactions between the most stable dimer systems and a graphene surface. 
All together, this dataset enables the exploration of key binding features (BFs) induced by BOV adsorption such as adsorption energy\cite{zhang2009_recovery_time}, charge transfer\cite{wehling2008_CT}, and work function change\cite{mathew2021schottky}.
This collection of BFs represent a big step towards the rational design and optimization of BOV–receptor systems due to the comprehensive electronic description of sensing performance; however, there are still some additional challenges to address before developing a sustainable framework for BOV–receptor design. 
For instance, analogous to ligand-pocket motifs \cite{Ryde2016,puleva2025}, the sensing process is inherently dynamic and governed by weak non-covalent interactions (electrostatics and hydrogen bonding), indicating a structural flexibility that yield a rugged energy landscape with myriad local minima and versatile binding configurations\cite{jeindl2022polymorphism}.
On the other hand, current theoretical models lack the quantitative rigor required to quantitatively delineate property–property and structure–property relationships, hindering a clear understanding of the role of sensing building blocks in BF behavior.

A promising approach to elucidate the complex mappings between atomic structures and BFs is the use of machine learning (ML) methods.
For instance, Ulissi \etal recently introduced the AdsorbML framework \cite{lan2023adsorbml_ads1}, which integrates heuristic search with ML potentials to accelerate gas–metal adsorption energy calculations, achieving both high predictive accuracy and substantial computational speedups compared to conventional density functional theory (DFT).
Similarly, GAME-Net \cite{pablo2023fast-gamenet}, a graph neural network model, was developed to predict adsorption energies of organic molecules on catalytic surfaces with near-DFT accuracy, reaching errors of 0.18 eV (0.016 eV per atom) for large biomass and plastic fragments.
More recently, Chen \etal introduced AdsMT \cite{chen2025multi_ads1}, a multimodal Transformer that combines catalyst surface graph representations with adsorbate feature vectors through a cross-attention mechanism to predict global minimum adsorption energies without enumerating adsorption sites.
While various ML-based studies \cite{tran2018_ads_10,zhong2020_ads11,fung2021_ads12,xu2022predicting_ads13,li2024interpreting_ads14} have focused on the adsorption of small and simple adsorbates (\eg \ce{O2}, \ce{CO2}, and \ce{H2}) on flat metal surfaces, other electronic BFs, such as charge transfer and work function change, have been less explored.
%
In addition, for large interacting molecules like the BOV-receptor systems, the concepts of binding site and adsorption distance become ill-defined owing to complex interaction morphologies and configurational polymorphism, which makes the development of predictive models more challenging. 
Furthermore, most of these works prioritize achieving high predictive accuracy, often at the expense of model interpretability, thereby limiting the physical and chemical insights that can be derived from these complex mappings.
This lack of explainability also affects the exploration of the binding features space, where the optimization of one feature offers no guarantee of concurrent improvements in others, complicating further the rational design of sensing materials.

\begin{figure}[t!]
    \centering
    \includegraphics[width=1\linewidth]{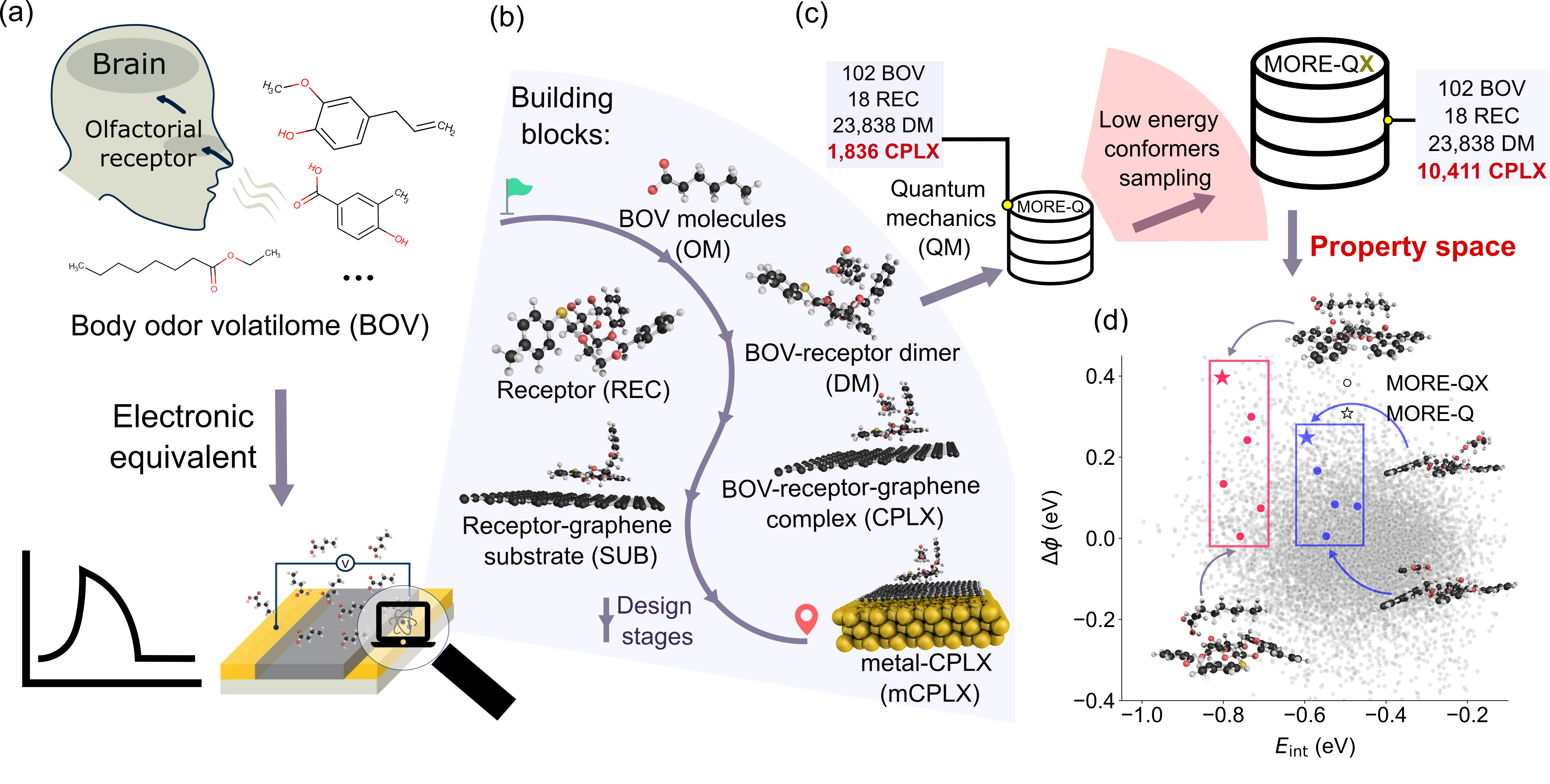}
    \caption{The schematic workflow for \textbf{M}olecular \textbf{O}lfactorial \textbf{R}eceptor \textbf{E}ngineering by \textbf{Q}uantum mechanics (MORE-Q)\cite{more-q} dataset expansion to MORE-QX dataset. (a) The bio-electronic noses (top right panel) are designed as an electronic equivalent to the olfactory system (top left panel), \eg for sensing BOV molecules (or odorant molecules, OM). (b) The building blocks at different design stages for the bio-mimetic sensor from QM perspective including OM molecules, molecular receptor (REC), OM-REC dimer molecule (DM), REC-graphene substrate (SUB), OM-REC-graphene complex system (CPLX) and eventually these systems deposited on the gold electrode (mCPLX). These abbreviations are used throughout this manuscript. 
    (c) The QM properties of the relevant building blocks were calculated and incorporated into the MORE-Q dataset, which includes monomer systems of 102 OM and 18 REC molecules, 23,838 DM systems, and 1,836 CPLX systems derived from the most stable DM configurations. Sampling multiple low-energy DM conformers expanded the CPLX subset, yielding the MORE-QX dataset with 10,411 CPLX systems.
    (d) 2D projection of the high-dimensional MORE-QX property space defined by the work function change $\Delta \phi$ and DM interaction energy $E_{\rm int}$. The conformers of two DM systems (red and blue) are labeled, where the most stable one (MORE-Q) is marked as star while the other low-energy conformers (MORE-QX) are marked with circles. The atomic structure associated to the maximal and minimal values are depicted on the plot for each DM system.}
    \label{FIG1}
\end{figure}

To address these challenges, we develop the MORE-ML framework, which integrates QM-derived molecular properties with ML methods to investigate how structural, global, and atomic-level features of electronic-nose building blocks influence BOV adsorption. 
By doing so, we seek to clarify the sensing mechanism and formulate design principles for BOV–receptor systems in artificial sensing materials.
To approximate the thermodynamic ensemble, we expanded the MORE-Q dataset\cite{more-q} into MORE-QX by sampling multiple low-energy BOV–receptor dimer (DM) conformers adsorbed on graphene. This process increased the number of BOV–receptor–graphene complexes from 1,836 to 10,441 (see Fig. \ref{FIG1}).
A comprehensive analysis of MORE-QX reveals that DM conformers with similar binding energies can nevertheless show markedly different BFs such as charge transfer and work function change.
Furthermore, DM properties and BFs exhibit only weak to moderate correlations, even though these properties were chosen following fundamental physical and chemical principles.
Despite particularly weak correlations among BFs, we retain flexibility in identifying systems that share a similar set of electronic binding characteristics---clear evidence for the existence of ``Freedom of design'' in the MORE-QX property space\cite{sandonas2023freedom}.
To enable rapid and accurate navigation of the binding feature space--and thereby support practical design of BOV–receptor complexes for sensing--we develop tree-based regression models that map QM-derived property data of building blocks to their associated BFs.
Within MORE-ML, we further exploit the interpretability of these models using SHapley Additive Explanations (SHAP)\cite{SHAP2017} to extract mechanistic insights into the sensing process.
Overall, this work provides quantum-informed understanding of adsorption mechanisms and enables the rational design of BOV–receptor systems, paving the way for controlled and robust discovery of artificial sensing materials.

\section*{Results}

\subsection*{Assessing conformational effects on binding features}

The intrinsic flexibility of large molecular receptors (REC) brings a crucial factor to consider in the understanding of the sensing mechanism in artificial olfactory sensors. 
Analogous to the binding process of ligands into protein pockets\cite{Ryde2016,puleva2025}, the interaction between odorant molecules (OM) and molecular receptors is inherently dynamic. Indeed, DM systems (refer to as dimer system) continually interconvert among multiple conformations, \ie jumping between minima on the potential energy surface (PES). 
Previous studies have shown that conformations of large molecules in both gas-phase\cite{aqm} and deposited on graphene nanoribbon\cite{Ravera2025} can exhibit comparable quantum-mechanical (QM) properties, raising the question of whether such effects also occur in these DM systems.
%
Therefore, instead of considering only the most stable conformer for each DM configuration, as is common in many DFT studies, we sampled a broader ensemble of low-energy conformers and adsorbed them onto graphene. This procedure expands the MORE-Q dataset \cite{more-q} into MORE-QX and increases the number of CPLX systems from $1,836$ to $10,441$ (see Figs. \ref{FIG1}(a-c)). Accordingly, the same set of QM properties computed for the CPLX systems in MORE-Q was also calculated for the additional DM conformations (see Methods section).
%
Note that the number of sampled conformers is adaptively adjusted to the morphological complexity of the DM system to ensure robust sampling. This means that systems with more flexible morphologies will yield a larger number of sampled conformers. On average, we considered six conformers per DM system. 
More details for the conformational sampling of DM systems can be found in Ref. \cite{more-q}.

A two-dimensional (2D) projection of the high-dimensional property space spanned by CPLX systems in MORE-QX is presented in Fig. \ref{FIG1}(d). Here, we illustrate the property space defined by the work function change ($\Delta \phi$) and the dimer interaction energy ($E_{\mathrm{int}}$), \ie $\left( \Delta \phi, E_{\mathrm{int}} \right)$.
Overall, one can see a lack of correlation between both properties, which indicates a degree of  flexibility when searching for dimer conformations with a given pair of $\left(\Delta \phi, E_{\mathrm{int}} \right)$ values.
To understand better the influence of conformational sampling, two example configurations were selected, see rectangles in Fig. \ref{FIG1}(d). In the red rectangle, $\Delta \phi$ is also uncorrelated with $E_{\mathrm{int}}$ and displays a large variation in magnitude with respect to the value corresponding to the most stable conformation, from $0.0$ to $0.4$ $\mathrm{eV}$. This change is also much larger compared to $E_{\mathrm{int}}$ that only decreases from $-0.8$ to $-0.75$ $\mathrm{eV}$ (\ie $\sim 0.05\,\mathrm{eV}$). 
Similarly, in the second set of studied conformations (enclosed by the rectangle blue), $E_{\mathrm{int}}$ is reduced because the OM molecule is changed by a smaller one, but $\Delta \phi$ still covers a larger property range ($\sim 0.25\,\mathrm{eV}$) 
This flexibility persists across the entire $\left( \Delta \phi, E_{\mathrm{int}} \right)$ property space, independent of the chosen DM configuration, underscoring the complexity of inferring binding features from QM properties of DM conformations.
%
Moreover, this result already indicates the challenge in determining simple physical and chemical rules for the simultaneous optimization of properties in the MORE-QX property space (\textit{vide infra}).
Nevertheless, Fig. \ref{FIG1}(d) conveys another important message for designing artificial olfactory systems: given a fixed $E_{\mathrm{int}}$, we might be able to find multiple CPLX systems with a desired $\Delta \phi$ value within a large range. Inversely, it is also possible to find different DM configurations with a desired $E_{\mathrm{int}}$ value in a large $\Delta \phi$ range. 
These initial observations provide the first evidence of an intrinsic ``Freedom of design'' in the MORE-QX property space\cite{sandonas2023freedom}, which will be discussed in the context of the binding feature space in the next section (see Fig. \ref{FOD}).
Additional property distributions representing the effect of conformational sampling can be found in Fig. S1 of the Supplementary Information (SI).

\begin{table}[t!]
    \centering
        \caption{List of relevant physicochemical properties for BOV-receptor (dimer systerm, DM) and BOV-receptor-graphene (complex system, CPLX) interaction. Each property presents a name, symbol, unit.  $a_0$ and D refer to the atomic unit of Bohr radius and Debye.}
    \begin{tabular}{cccc}
    \toprule
       \#& Property & Symbol & Unit \\
       \hline
        1 & Interaction energy& $E_{\mathrm{int}}$ & eV \\
        2 & Isotropic molecular polarizability & $\alpha_{\rm s,DM}$ & $a_0^{3}$ \\
        3 & Scalar dipole moment & $\mu_{\rm DM}$ & D \\
        4 & Dipole moment component along slab ($z$) direction  & $\mu_{z,\rm DM}$ & D \\
        5 & HOMO energy & $\epsilon_{\rm H,DM}$ & eV \\
        6 & LUMO energy & $\epsilon_{\rm L,DM}$ & eV \\
        7 & HOMO-LUMO gap & $\epsilon_{\rm gap}$ & eV \\
        8 & Adsorption eneregy & $\epsilon_{\rm gap}$ & eV \\
        9 & Work function change & $\Delta \phi$ & eV \\ 
        10 & Charge transfer & $\Delta Q$ & e \\ 
        
        \hline
    \end{tabular}

    \label{ORCA-table}
\end{table}

\subsection*{``Freedom of design'' in the MORE-QX property space}

\begin{figure}[!ht]
    \centering
    \includegraphics[width=0.5\linewidth]{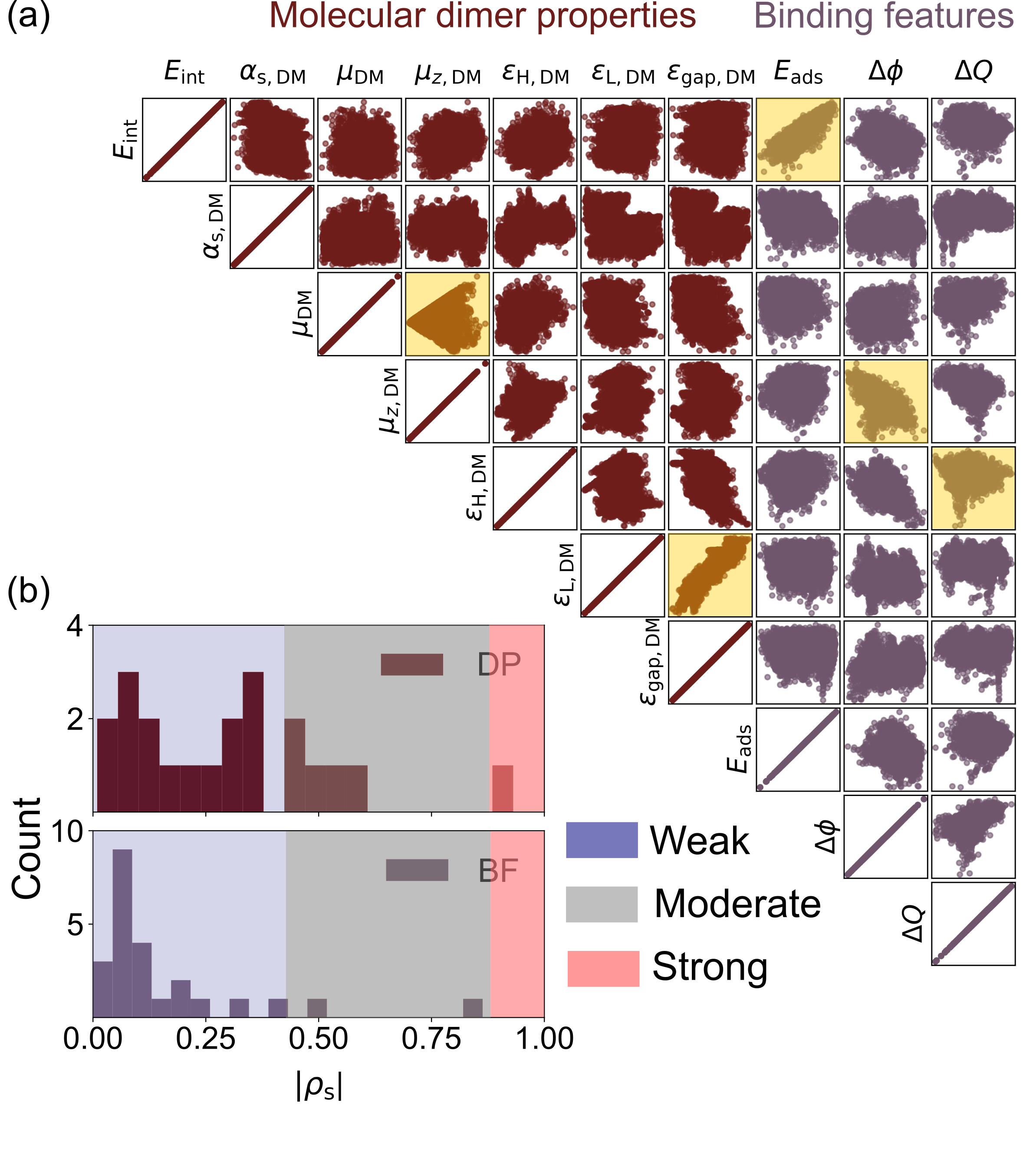}
    \caption{(a) Two-dimensional (2D) projections of the high-dimensional property space spanned by MORE-QX dataset. We show the correlation plots for seven dimer properties (DP, brown) and three binding features (BF, purple) from MORE-QX. The detailed description of the properties can be viewed in Table \ref{ORCA-table}. Some interesting projections are marked by yellow frames and discussed in the manuscript. (b) The count measurement of absolute value of Spearman correlation coefficient $|\rho_s|$ for DP (upper panel) and BF (lower panel) 2D projections. The $|\rho_s|$ values result in three distinct clusters: weakly correlated $|\rho_s| \leq 0.5$, moderately correlated $0.5 < |\rho_s| \leq 0.8$, and strongly correlated $|\rho_s| > 0.8$ covering by blue, gray, red frames, respectively.
    }
    \label{FOD}
\end{figure}

To gain a deeper understanding of the relationship between the QM properties of the building blocks and the resulting binding features (BFs), we examined selected pairwise correlations within the high-dimensional property space spanned by MORE-QX.
Specifically, we analyzed correlations between the properties of DM systems and the associated BFs (see the full property list in Table~\ref{ORCA-table}).
DM properties were selected because of their strong involvement in physicochemical effects arising from molecule–surface interactions, such as orbital hybridization, polarization effects, charge density redistribution, and charge transfer, which ultimately influence the binding features~\cite{DFT_ads_process_2011}.  
Overall, Fig.~\ref{FOD}(a) shows that nearly all of the 45 unique pairwise projections (\ie 2D correlation plots) resemble structureless ``blobs'', indicating that most of these QM properties are effectively uncorrelated.
To quantify the degree of correlation, we computed the absolute value of the Spearman correlation coefficient, $|\rho_s|$ (see Eq.~\ref{spearmann}).
The distributions of $|\rho_s|$ for DM properties and BFs are shown in the upper and lower panels of Fig.~\ref{FOD}(b), respectively, where the pairwise correlations are categorized according to their $|\rho_s|$ values.
Properties are considered strongly correlated if $|\rho_s| > 0.8$, moderately correlated if $0.5 < |\rho_s| \leq 0.8$, and weakly correlated if $|\rho_s| \leq 0.5$.
Accordingly, among the DM properties, 1 out of 21 pairwise correlations ($\approx 4.8\%$) is strongly correlated, 4 out of 21 ($\approx 19\%$) are moderately correlated, and the remaining 16 ($\approx 76\%$) are weakly correlated. 
In contrast, none of the 24 correlations associated with BFs are strongly correlated; 2 out of 24 ($\approx 8\%$) exhibit moderate correlation, while the remaining 22 ($\approx 92\%$) are weakly correlated.
This comparison demonstrates that correlations are generally weak for both DM properties and BFs, with correlations among BFs being even weaker than those among DM properties. This behavior reflects a more intricate and nontrivial interplay of interatomic interactions in OM–REC–graphene (CPLX) systems compared to single dimers (OM-REC systems).

Among the 2D property spaces analyzed, a few cases of interest exhibit moderate to strong correlations (highlighted by yellow frames in Fig. \ref{FOD}(a)).
For example, the HOMO-LUMO gap ($\epsilon_{\mathrm{gap,DM}}$) of DM systems shows a more linear correlation with the LUMO energy ($\epsilon_{\mathrm{L,DM}}$) than with the HOMO energy ($\epsilon_{\mathrm{H,DM}}$). 
This observation implies that $\epsilon_{\mathrm{H,DM}}$ can be used to distinguish DM systems with similar $\epsilon_{\mathrm{gap,DM}}$, which is an important requirement for constructing efficient electronic descriptors.
%
Regarding correlations with BFs, the interaction energy ($E_{\mathrm{int}}$) of DM systems and the corresponding adsorption energy ($E_{\mathrm{ads}}$) exhibit a strong correlation, with $|\rho_s| = 0.86$. 
This result suggests that the interaction mechanism between OM and REC systems can be transferred to CPLX systems to describe trends in $E_{\mathrm{ads}}$. 
However, the correlation is not fully linear, indicating that fluctuations arise from geometry and charge-distribution changes induced by surface interactions.
Another relevant BF is the adsorbate-induced charge transfer ($\Delta Q$), which is commonly interpreted within the orbital-mixing theory that describes the alignment between the substrate Fermi level (the DM system in this work) and the frontier orbital energies of the adsorbate \cite{2015_orbital_mixing}. 
By computing $|\rho_s|$ between $\Delta Q$ and the orbital energies of DM systems, we find that both HOMO and LUMO energies are only weakly correlated with $\Delta Q$, with $|\rho_s| = 0.02$ and $|\rho_s| = 0.11$, respectively. 
Similarly, orbital energies associated to OM systems are also uncorrelated with $\Delta Q$, yielding $|\rho_s| < 0.3$.
This lack of correlation reveals the complexity of using orbital energies alone to define design principles for tuning $\Delta Q$. At the same time, it reflects a certain ``freedom of design'' within the binding feature space, enabling the identification of DM systems with targeted orbital energies that can serve as components of electronic descriptors for BF prediction (\textit{vide infra}).
%

In our correlation analysis with the work function change ($\Delta \phi$), we found that the z-component of the dipole moment in the DM ($\mu_{\mathrm{z,DM}}$) and OM ($\mu_{\mathrm{z,OM}}$) systems shows a moderate correlation with $\Delta \phi$, with $|\rho_s| = 0.51$ and $|\rho_s| = 0.68$, respectively.
As discussed in our previous work \cite{DWF_DIP_Li,DWF_mohannmad,DWF_leung}, $\Delta \phi$ follows the Helmholtz relation:
\begin{equation}
\Delta \phi = - e/\varepsilon_{\mathrm{0}}\cdot \Delta P_{\mathrm{tot}}, 
\label{surface_DIP_1}
\end{equation}
where surface dipole moment change ($\Delta P_{\mathrm{tot}}$) could be split into several components as,
\begin{equation}
    \Delta \phi= - e/\varepsilon_{\rm0} \cdot (\Delta p_{\mathrm {cplx}} + p_{\mathrm {a}} + p_{\mathrm{s}} - p_0 ),
    \label{surface_DIP_2}
\end{equation}
where the components $\Delta p_{\mathrm {cplx}}$, $p_{\mathrm {a}}$, and $p_{\mathrm{s}} - p_0$ denote the adsorbate-induced surface dipole moment change by spatial charge redistribution, adsorbate dipole moment, and surface deformation, respectively. 
$\mu_{\mathrm{z,DM}}$ inherently contains information related to $\mu_{\mathrm{z,OM}}$, which is tightly associated with the $p_a$ term and yields a moderate correlation ($|\rho_s| = 0.51$).
However, other contributions---particularly $\Delta p_{\mathrm{cplx}}$, which describes spatial charge redistribution---are poorly captured by $\mu_{\mathrm{z,DM}}$ or by any other DM property \eg polarizability $\alpha_{\rm S,DM}$, as evidenced by the very low correlation ($|\rho_s| = 0.07$ between $\Delta \phi$ and $\alpha_{\rm S,DM}$).
These findings highlight both the intrinsic complexity of $\Delta \phi$ and the insufficiency of current physicochemical heuristics for tailoring it.
%
While the weak-to-moderate correlations between DM properties and BFs provide some theoretical guidance based on physicochemical intuition, no clear patterns emerge to navigate the binding feature space.
Moreover, there is little correlation among the BFs themselves. For example, $E_{\rm ads}$ is only weakly correlated with $\Delta \phi$ ($|\rho_s| = 0.14$). Likewise, $\Delta Q$ shows a weak correlation with $E_{\rm ads}$ ($|\rho_s| = 0.05$) and a moderate correlation with $\Delta \phi$ ($|\rho_s| = 0.40$), the latter arising from spatial charge redistribution upon adsorption \cite{DWF_leung,DWF_DIP_Li}.
Collectively, these observations indicate that only few constraints limit a DM system from simultaneously exhibiting any given pair of DM and BF properties considered in Fig.~\ref{FOD}(a), providing compelling evidence for the existence of a ``freedom of design'' in the binding feature space.
Building on this concept, we also analyzed how the weak correlations among BFs enable the identification of DM conformations tailored to specific target properties (see the SI for details).
%
Consequently, a large number of electronic features may serve as efficient molecular descriptors for BF prediction; however, owing to differences in correlation strength and underlying physicochemical insight, some descriptors are likely to be more relevant than others. 

Notice that, even though Boltzmann-weighted properties could in principle be used to construct more accurate ensembles \cite{kim2025functional}, we treat each low-energy dimer conformer equally in order to probe conformer-specific effects and to explore the potential energy surface more comprehensively than a static Boltzmann average would allow.
Because the low-energy conformers have similar Boltzmann weights, weighting or direct averaging would obscure subtle inter-conformer differences. Since our primary goal is to examine how BFs vary across individual surface-bound dimers, we therefore do not apply Boltzmann weighting.

\begin{figure}[t!]
    \centering
    \includegraphics[width=1\linewidth]{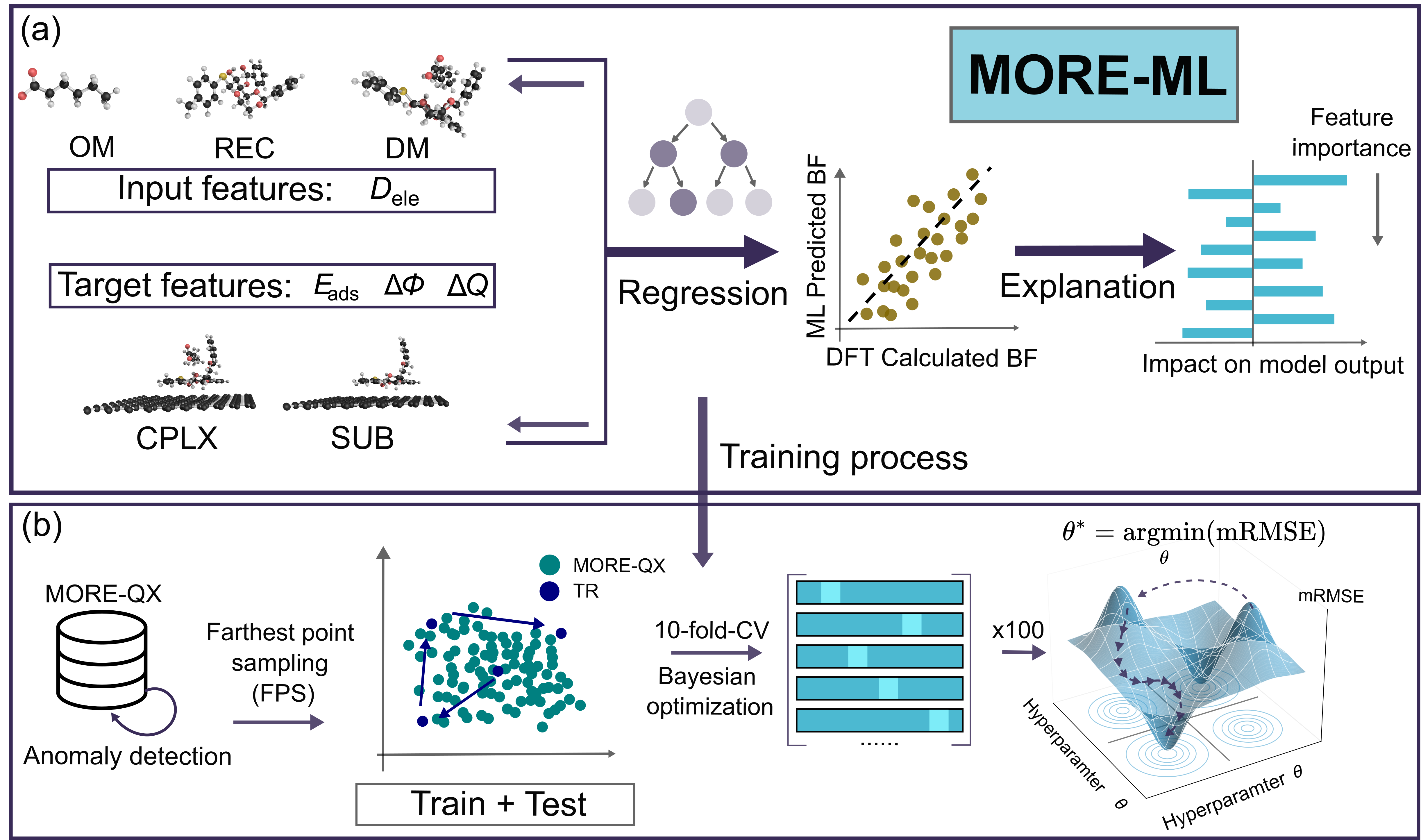}
    \caption{(a) Scheme of the MORE-ML framework, which stands for \textbf{M}olecular \textbf{O}lfactorial \textbf{R}eceptor \textbf{E}ngineering by \textbf{M}achine \textbf{L}earning, which integrates QM properties of molecular building blocks ($\rm D_{ele}$) with ML techniques for the prediction of binding features (BFs) such as $E_{\rm ads}$, $\Delta \phi$ and $\Delta Q$.
    MORE-ML framework aims at regression and model explanation tasks. (b) ML model training in MORE-ML starts with anomaly detection (see SI), followed by farthest point sampling (see Methods) to construct the training and test sets. Bayesian optimization with 100 iterations and 10-fold cross-validation on the training set is used for hyperparameter tuning. Final model performance is evaluated on the test set.
    }
    \label{ML_workflow}
\end{figure}

\subsection*{Navigating the binding feature space via machine learning}

Although the lack of correlation among BFs provides a flexibility in designing CPLX systems with desired sensing-related properties, determining the relationship between BFs and electronic properties of molecular building blocks (OM, REC, and DM systems) is still challenging. 
To address this issue, we have implemented the machine learning (ML) framework MORE-ML (see Fig. \ref{ML_workflow}), which aims at establishing a quantitative and explainable mapping between these property spaces by using ML regression techniques and explainable AI methods (see Methods). 
%
%
To identify the most suitable regression models for BF prediction, we benchmark the performance of several tree-based methods: random forest (RF) \cite{breiman2001RF}, gradient boosting decision trees (GB), XGBoost (XGB) \cite{chen2016xgboost}, CatBoost (CAT) \cite{prokhorenkova2018catboost}, and LightGBM (LGBM) \cite{ke2017lightgbm}.
The best-performing models will be subsequently analyzed using SHapley Additive exPlanations (SHAP) \cite{SHAP2017}, an efficient explainable AI framework well suited to tree-based models.
As a training strategy, we prioritized electronic-structure–derived descriptors ($D_{\rm ele}$), composed of QM properties of OM, REC, and DM systems, owing to their lightweight nature and clear physicochemical interpretability (see Table S3 in the SI).
Moreover, inspired by the development of the QUED framework \cite{hinostroza25}, we investigated whether model performance could be further improved by combining $D_{\rm ele}$ with geometrical descriptors $D_{\rm geo}$ (\eg Bag-of-Bonds\cite{hansen2015bob}, SOAP\cite{sandip2016soap}, and MACE\cite{kovacs2025mace}) and the corresponding Mulliken atomic charges $q$ ($D_q$).
However, as shown in Figs. S7 and S8 of the SI, the inclusion of these additional descriptors did not improve the performance of the ML models. This lack of improvement may be attributed to redundant geometrical information arising from the presence of similar OM and REC systems across multiple DM structures (\textit{vide supra}).
%
%
Accordingly, we performed a more in-depth analysis of ML model accuracy using only $D_{\rm ele}$.

\begin{figure}[t!]
    \centering
    \includegraphics[width=0.65\linewidth]{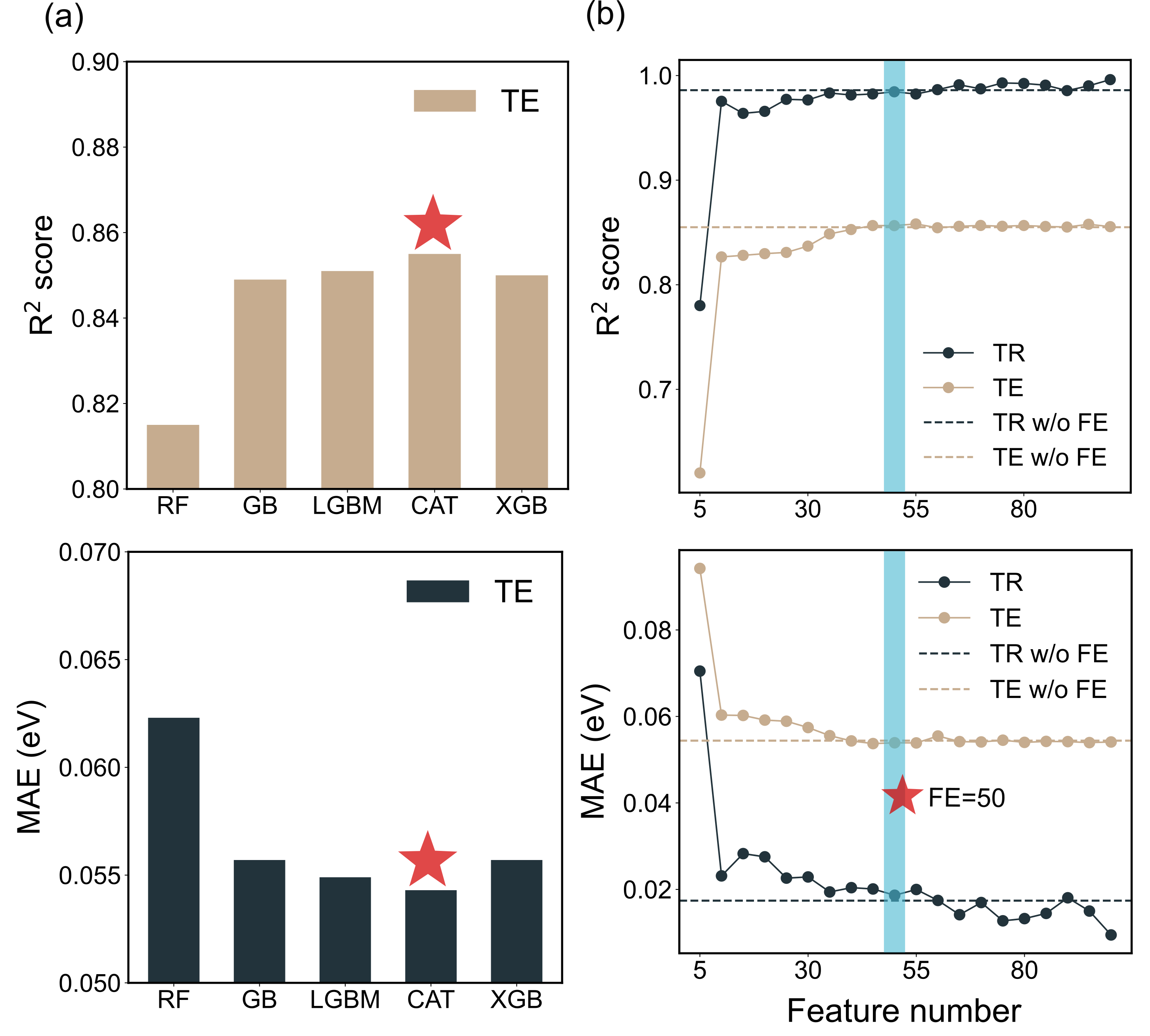}
    \caption{Model benchmarking and feature engineering for the prediction of adsorption energy ($E_{\rm ads}$). (a) Coefficient of determination ($R^2$) and mean absolute error (MAE) evaluated on the test (TE) set for tree-based models: random forest (RF), gradient boosting decision tree (GB), LightGBM (LGBM), CatBoost (CAT), and XGBoost (XGB). The best-performing model is indicated by red stars. (b) Evolution of $R^2$ (upper panel) and MAE (lower panel) during feature engineering (FE) of the CatBoost model for predicting $E_{\rm ads}$. The number of features is increased incrementally in steps of five, ranked by SHAP analysis (see Methods). Model performance is shown for the training (TR, black) and test (TE, brown) sets. Dashed lines indicate performance obtained using the full set of QM properties as features.}
    \label{benchmark&FE}
\end{figure}

Fig. \ref{benchmark&FE}(a) shows the coefficient of determination ($R^2$) and mean absolute error (MAE) for predicting the adsorption energy, $E_{\mathrm{ads}}$, using the full $D_{\rm ele}$ descriptor (130 features).
%
The MORE-QX dataset was partitioned into training (TR) and test (TE) sets using a fixed 9:1 ratio. Additional details on the dataset splitting procedure and the selection of training samples are provided in the Methods section.
By comparing the results obtained for $E_{\mathrm{ads}}$ with those corresponding to other binding features (see Fig. S7 in the SI), we find that all benchmarked methods exhibit similar performance trends across features. Consequently, $E_{\mathrm{ads}}$ is used here as a representative case.
Among the tree-based methods, RF performs the worst in the TE set, yielding an $R^2$ of 0.818 and an MAE of 0.062 eV. This suggests that gradient-boosting approaches outperform RF's bagging strategy in capturing latent correlations between $D_{\rm ele}$ and the binding features. A likely explanation is that, in boosting, each successive tree corrects the residuals of its predecessor, whereas RF relies on an ensemble of independent trees.
%
Within the gradient-boosting family, CAT achieves the best performance, with an $R^2$ of 0.86 and an MAE of 0.058 eV. This advantage likely arises from the use of oblivious (symmetric) trees, in which all nodes at a given depth split on the same feature.
Such a structure imposes strong regularization on tree complexity, thereby improving generalization in binding feature prediction.
As a result, we adopt CAT as the final ML regression model for subsequent analyses.

Then, we focus on selecting the most informative subset of QM properties within $D_{\rm ele}$ to mitigate high dimensionality and reduce model noise. 
By identifying and removing redundant and highly correlated features, we aim to prevent overfitting and improve the generalizability of the ML model.
To this end, we employed an iterative SHAP-driven feature selection procedure using the CAT models. At each iteration, the model is retrained with a reduced subset of the full $D_{\rm ele}$, consisting of the top-ranked features according to SHAP importance.
The number of selected features was gradually increased from 5 to 105 in increments of 5, and model performance was evaluated at each stage.  The resulting learning curves for the $R^2$ and MAE metrics are shown in Fig.~\ref{benchmark&FE}(b).
Based on the results for the TR and TE sets, the feature learning behavior can be divided into growing and saturated regimes.
In the small-$D_{\rm ele}$ regime, performance improves gradually but remains inferior to that achieved with the full descriptor set (see dashed lines), indicating that an insufficient number of QM properties is available to accurately capture the adsorption mechanism.
Once the size of $D_{\rm ele}$ exceeds a critical threshold, the performance curves begin to plateau: the TE scores no longer improve, while the TR scores show only minor fluctuations.
This behavior indicates that additional features do not further enhance the model’s understanding of the adsorption mechanism, suggesting the existence of an optimal QM subset that balances accuracy and efficiency.
Based on this exhaustive analysis, we selected the top 50 electronic features (star-labeled) as the effective descriptor set for $E_{\mathrm{ads}}$.
Using the same procedure, the top 60 and top 80 features were selected for $\Delta \phi$ and $\Delta Q$, respectively (see Fig. S7 in the SI).

\begin{figure}[t!]
    \centering
    \includegraphics[width=1\linewidth]{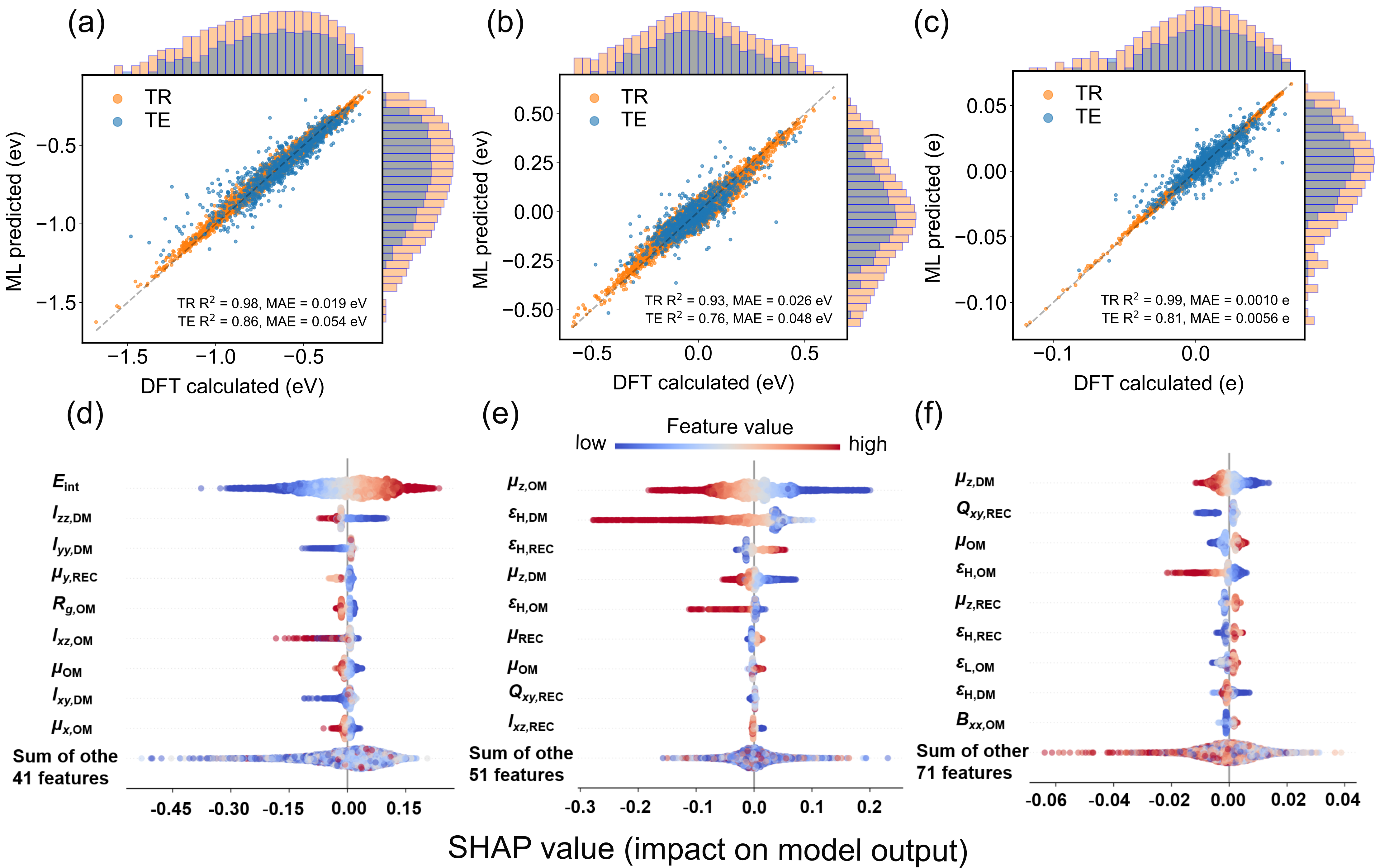}
    \caption{Correlation plots between DFT calculated and ML predicted values are shown for the best-performing models used to predict (a) $E_{\rm ads}$, (b) $\Delta \phi$, and (c) $\Delta Q$. Orange and blue bars/points represent the training (TR) and test (TE) sets, respectively. The lateral panels display the distributions for each binding feature.
   Panels (d–f) show the corresponding SHAP beeswarm plots (see Methods) for (d) $E_{\rm ads}$, (e) $\Delta \phi$, and (f) $\Delta Q$. In each beeswarm plot, features are ranked in ascending order of importance from top to bottom, with SHAP values distributed around the zero baseline. Each point is colored according to the corresponding feature value. Only the top nine features are shown; the cumulative SHAP value of all remaining features is reported in the final column ($10^{\rm th}$ position).}
    \label{ML_results_SHAP}
\end{figure}

Indeed, the final ML regression models for predicting $E_{\rm ads}$, $\Delta \phi$ and $\Delta Q$ were developed using the optimized subset of QM features and CAT method (see Fig. \ref{ML_results_SHAP}).
%
To assess their overall learning capability, we first examine the $\rm R^2$  metric, which quantifies the variance between DFT-calculated and ML-predicted values.
For TR set, $\rm R^2$ reaches $0.99$ for both $E_{\rm ads}$ and $\Delta Q$, whereas a slightly lower value of $0.93$ is obtained for $\Delta \phi$.
This difference is reflected in the larger dispersion of the orange data points around the $y=x$ reference line (dashed).
Considering the MAE metric, the corresponding values for TR set of $E_{\rm ads}$ and $\Delta Q$ are $0.017\,\rm eV$, and $0.001\,\rm e$, respectively, while $\Delta \phi$ exhibits a higher MAE of $0.026\,\rm eV$. 
Given the discrete nature of the binding feature space and the limited coverage of MORE-QX dataset, we further evaluate model performance using the relative error  $\epsilon = \dfrac{|y_{\rm ML} -y_{\rm DFT}|}{\Delta y}\times100$ with $y_{\rm ML}$ and $y_{\rm DFT}$ as the ML and DFT values of the property $y$.
$\Delta y$ represents the extent of the property spectrum across the entire dataset. 
The resulting relative errors for $E_{\rm ads}$, $\Delta \phi$ and $\Delta Q$ are $1.7\%$, $2.6\%$ and $1\%$, respectively. These small values indicate that the models accurately reproduce the training data.

We next examine the generalization capability of the ML models by evaluating their performance on unseen systems considered in the TE set.
As expected, model accuracy decreases relative to the TR set, yielding $\rm R^2$ values of $0.86$, $0.76$, and $0.81$ for $E_{\rm ads}$, $\Delta \phi$, and $\Delta Q$, respectively.
The relative errors also increases, but remain below $6\%$: $5.4\%$ for $E_{\rm ads}$, $4.8\%$ for $\Delta \phi$, and $5.6\%$ for $\Delta Q$.
The moderate performance gap between the TR and TE sets indicates that the models retain high predictive accuracy for novel systems, underscoring their potential to generalize across a much larger configuration and conformational space.
This conclusion is further supported by the close agreement between the distributions of predicted binding features for the TR and TE sets (see right panels in Figs. \ref{ML_results_SHAP}(a-c)).
The complete set of evaluation metrics for each model is summarized in Table S6.
To elucidate the slightly reduced accuracy of the model predicting $\Delta \phi$, we separately predicted $\phi$ values for both the complex ($\phi_{\rm CPLX}$) and the substrate ($\phi_{\rm SUB}$) systems using the same TR and TE sets.  
As shown in Fig. S10, the predictions for $\phi_{\rm CPLX}$ yield $\rm R^2 = 0.87$ and $\mathrm{MAE} = 0.045\, \rm eV$; while those for $\phi_{\rm SUB}$ achieve $\rm R^2 = 0.89$ and $\mathrm{MAE} = 0.029\, \rm eV$.
Despite these favorable metrics, the parity plot for $\phi_{\rm SUB}$ exhibits an unexpected zigzag pattern in both the TR and TE sets, and the model fails to reproduce the bimodal distribution of $\phi_{\rm SUB}$. 
This shortcoming likely stems from the limited diversity of $\phi_{\rm SUB}$ values in the dataset: the QM descriptors of the building blocks, particularly those derived from the 18 REC structures, are insufficient to capture the subtle conformational variations that govern $\phi_{\rm SUB}$. Consequently, noise is introduced into the prediction of $\phi_{\rm SUB}$, even though $\phi_{\rm CPLX}$ is modeled accurately.

\subsection*{AI-based explanation of binding feature predictive models}
To better interpret the tree-based ML models developed for BF prediction, we performed an explainability analysis using both their intrinsic interpretability and SHAP method (see Methods).
%
The beeswarm plots in Figs. \ref{ML_results_SHAP}(d-f) summarize the distribution of SHAP values for the most influential features in each prediction task.
In these plots, features are ranked by importance from top to bottom, and their corresponding SHAP values are shown along the $x$-axis. 
Positive SHAP values indicate that a feature increases the predicted outcome, whereas negative values indicate a decrease.
The color gradient encodes the feature magnitude, with red representing high values and blue representing low values.

The SHAP value distribution in Fig. \ref{ML_results_SHAP}(d) clearly shows that the dimer (DM) interaction energy, $E_{\rm int}$, plays the most dominant role in determining $E_{\rm ads}$.
The color gradient indicates that smaller $E_{\rm int}$ values lead to smaller $E_{\rm ads}$ values and vice versa, since both quantities are negatives. 
This strong coupling between $E_{\rm int}$ and its SHAP value is reflected in the high Spearman correlation coefficient, $|\rho_s| = 0.86$, indicating that $E_{\rm int}$ serves as an effective descriptor for $E_{\rm ads}$ on the graphene surface. 
Although $E_{\rm int}$ contains the majority of the predictive information for $E_{\rm ads}$, the model still needs to account for a small residual difference between these two energetics to achieve higher accuracy.
This difference is captured by morphological descriptors, such as the components of the inertia tensor ($I$) of the DM systems and the radius of gyration ($R_g$) of the OM system, highlighting that molecular structure also plays a critical role. 
Furthermore, the dipole moments ($\mu$) of OM and REC systems rank among the top ten features, indicating that charge redistribution is relevant for describing non-covalent interactions during adsorption. 
The SHAP values of these additional features are distributed much more narrowly than those of $E_{\rm int}$, which explains their lower overall importance.
Consequently, these features-together with the remaining descriptors, primarily act as fine-tuning factors, capturing a small number of outliers and subtle corrections compared to the dominant contribution of $E_{\rm int}$.

\begin{figure}[t]
    \centering
    \includegraphics[width=1\linewidth]{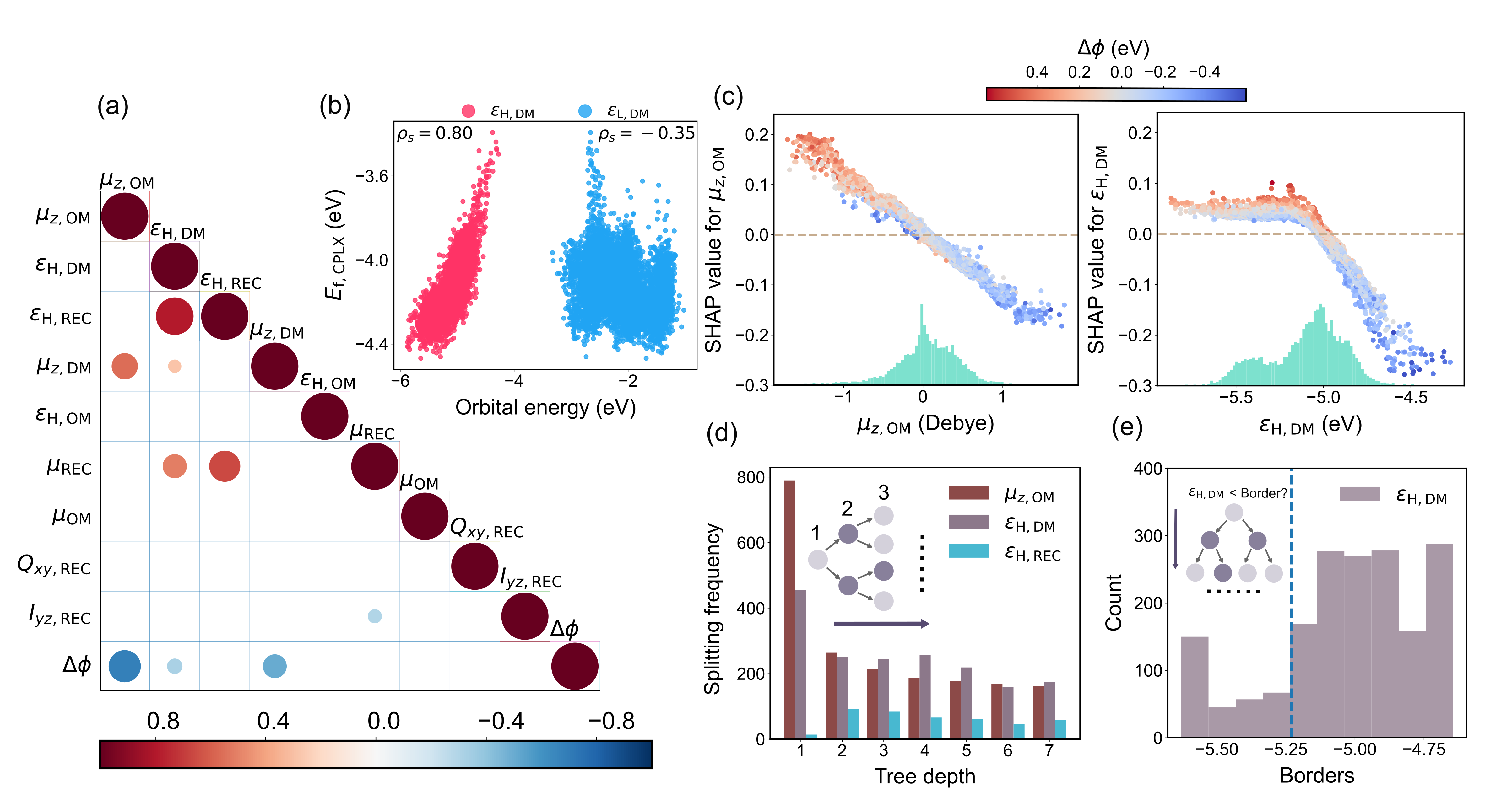}
    \caption{Explanation of the ML model for $\Delta \phi$ prediction. (a) Pairwise Spearman correlation coefficients $\rho_s$ between the top nine features and $\Delta \phi$. Circle size and color indicate the magnitude and sign of $\rho_s$, respectively. (b) Correlation plots between the Fermi level of the CPLX systems, $E_{\rm f, CPLX}$, and the dimer HOMO $\epsilon_{\rm H,DM}$ (red) and LUMO $\epsilon_{\rm L,DM}$ (blue) energies. Corresponding $\rho_s$ values are shown in the plots.
    (c) SHAP value dependence plots for the two most important features: vertical dipole moment of OM $\mu_{\rm z,OM}$ (left panel) and dimer HOMO energy $\epsilon_{\rm H,DM}$ (right panel). SHAP values are shown as a function of the corresponding feature value; data points are colored by $\Delta \phi$, with feature value distributions shown along the $x$-axis. 
    (d) Frequency of feature participation in node splits as a function of tree depth for the three most important features: $\mu_{\rm z,OM}$, $\epsilon_{\rm H,DM}$, and the receptor HOMO energies $\epsilon_{\rm H,REC}$. (e) Distribution of splitting frequencies as a function of the border values for $\epsilon_{\mathrm{H,DM}}$.}
    \label{explanation}
\end{figure}

A similar trend can be observed in the SHAP analysis for predicting $\Delta Q$ (see Fig.~\ref{ML_results_SHAP}(f)).
In contrast to the morphology-correlated binding feature $E_{\rm ads}$, $\Delta Q$ is primarily correlated with charge-related properties.
In particular, the dipole and quadrupole moments emerge as the most relevant features, whereas molecular orbital energies appear lower in the ranking; \eg $\epsilon_{\rm H,\rm OM}$ and $\epsilon_{\rm H,\rm REC}$ occupy the $4^{\rm th}$ and $6^{\rm th}$ positions, respectively.
This indicates that several properties contribute synergistically to the prediction of $\Delta Q$, with no single dominant feature.
Moreover, the SHAP analysis highlights the limited capability of a purely qualitative orbital-mixing description of charge transfer \cite{2015_orbital_mixing}, as the frontier orbital energies are not among the dominant predictors.
%
Unlike $E_{\rm ads}$ and $\Delta Q$, the prediction of $\Delta \phi$ is governed by two dominant features: the vertical dipole moment of the OM, $\mu_{z,\rm OM}$, and the HOMO energy of the DM system, $\epsilon_{\rm H,\rm DM}$, which exhibit a wide distribution in Fig. \ref{ML_results_SHAP}(e). 
The importance of $\mu_{z,\rm OM}$ is readily explained by Eq.~\ref{surface_DIP_2}, since it directly contributes to the total change in the surface dipole moment.
Interestingly, the SHAP distribution of $\epsilon_{\rm H,\rm OM}$ (ranked $5^{\rm th}$) shows a trend similar to that of $\epsilon_{\rm H,\rm DM}$, which is the second most important feature.
A qualitative explanation for the high ranking of frontier orbital energies (HOMO/LUMO) is that $\Delta \phi$ partially originates from spatial charge redistribution. In this context, the HOMO and LUMO energies represent the primary donor and acceptor orbitals, respectively, thereby inducing charge-density changes on and near the associated atoms.

To gain further physical insight into the prediction of $\Delta \phi$, we first analyzed the Spearman correlation coefficient, $|\rho_s|$, between the top 10 QM features and $\Delta \phi$ (see Fig.~\ref{explanation}(a)).
Among these features, only a few properties exhibit clear correlations. For example, $\mu_{z,\mathrm{OM}}$ and $\mu_{z,\mathrm{DM}}$ are strongly correlated, as $\mu_{z,\mathrm{DM}}$ contains information from $\mu_{z,\mathrm{OM}}$. These features are also correlated with $\Delta \phi$ because they partially enter Eq.~\ref{surface_DIP_2}.
In contrast, the majority of the top 10 features show weak correlations ($|\rho_s| < 0.05$), indicating that SHAP-based feature ranking effectively mitigates multicollinearity among the QM descriptors. This procedure filters out highly correlated and thus noisy features, ultimately leading to improved generalization by leveraging a diverse set of non-redundant descriptors.  
%
Moreover, the counterpart $\epsilon_{\mathrm{L,DM}}$ of $\epsilon_{\mathrm{H,DM}}$ does not appear among the top 10 features, whereas $\epsilon_{\mathrm{H,DM}}$ ranks second. This suggests that the surface Fermi level predominantly interacts with $\epsilon_{\mathrm{H,DM}}$, consistent with orbital mixing theory \cite{2015_orbital_mixing}.
Consequently, $\epsilon_{\mathrm{H,DM}}$ tends to align with the surface Fermi level, and the resulting Fermi level of the complex system, $E_{\mathrm{f,CPLX}}$, is more strongly associated with $\epsilon_{\mathrm{H,DM}}$ than with $\epsilon_{\mathrm{L,DM}}$, with $\rho_s = 0.8$ and $\rho_s = -0.35$, respectively (see Fig.~\ref{explanation}(b)).
To further investigate the synergistic mechanisms of the most important QM features, \eg $u_{z,\rm OM}$ and $\epsilon_{\rm H,\rm DM}$, in tuning $\Delta \phi$, we analyze their contribution behavior by correlating SHAP values with property distributions (see Fig. \ref{explanation}(c)).
In the left panel, the SHAP values of $\mu_{\rm z,\rm OM}$ exhibit a clear linear correlation with the feature itself: negative $\mu_{\rm z,\rm OM}$ values yield positive contributions to $\Delta\phi$, and vice versa, with the sign determined by the direction of the surface dipole moment.
In general, larger absolute values of $\mu_{\rm z,\rm OM}$ lead to stronger contributions to $\Delta\phi$, consistent with its $\rho_s$ value.
In contrast, the SHAP values of $\epsilon_{\rm H,\rm DM}$ in the right panel remain nearly constant as $\epsilon_{\rm H,\rm DM}$ increases from its minimum up to approximately  $-5.25\,\rm eV$. Beyond this turning point, a linear correlation between SHAP values and $\epsilon_{\rm H,\rm DM}$ emerges as the feature value increases further.
%

These behaviors can be understood through the intrinsic interpretability of tree-based models. Owing to the hierarchical splitting process, features used at shallower tree depths acquire greater importance than those applied deeper in the tree, since early splits typically yield larger information gains by partitioning a larger fraction of the dataset.
As shown in Fig. \ref{explanation}(d), we quantify the frequency with which  $\mu_{z,\rm OM}$, $\epsilon_{\rm H, DM}$ and $\epsilon_{\rm H, REC}$ (ranked $4^{\rm th}$ in Fig. \ref{ML_results_SHAP}) participate in splits at each tree depth. 
At the first tree level, the bars corresponding to  $\mu_{z,\rm OM}$ and $\epsilon_{\rm H, DM}$ are markedly higher than that of $\epsilon_{\rm H, REC}$, with $\mu_{z,\rm OM}$ also significantly exceeding $\epsilon_{\rm H, DM}$.
At greater depths, $\mu_{z,\rm OM}$ and $\epsilon_{\rm H, DM}$ continue to participate frequently in splits, albeit with reduced information gain due to the smaller number of remaining data points.
Notably, $\epsilon_{\rm H,\rm DM}$ slightly surpasses $\mu_{z,\rm OM}$ in splitting frequency at deeper levels, corresponding to splits that isolate a small number of exceptional outliers. This compensates for the stronger early contribution of $\mu_{z,\rm OM}$, resulting in comparable total SHAP contributions for the two features, as reflected in the importance ranking in Fig. \ref{ML_results_SHAP}.
Overall, these observations confirm the dominant and widespread importance of $\mu_{z,\rm OM}$ and $\epsilon_{\rm H, DM}$, with $\mu_{z,\rm OM}$ retaining a slightly higher overall ranking. The shorter bar heights of $\epsilon_{\rm H,REC}$  are also consistent with its narrowly distributed SHAP values and, consequently, its lower importance.
Moreovero, the turning point at $\epsilon_{\rm H,\rm DM} \approx -5.25\,\rm eV$ can be illustrated by the participation of the property values (corresponding to decision borders in tree-based models) at the splitting nodes (see Fig. \ref{explanation}(e)).
The number of borders with values $> -5.25 \, \rm eV$ is significantly higher than those $\leq -5.25 \, \rm eV$, indicating that the values above this threshold appear more frequently in the split decisions.
In this regime, the property values are more continuous and lead to a broader range of contribution values, whereas borders $\leq -5.25 \, \rm eV$ participate much less frequently in the splitting process.
Indeed, the cumulative information gain from borders  $\leq -5.25 \, \rm eV$ results in only minor contributions, fluctuating between $0$ and $0.1\, \rm eV$  to $\Delta \phi$.
Contrarily, border $> -5.25 \, \rm eV$ yield large contribution with a broad distribution, consistent with the threshold effect shown in the left panel of Fig. \ref{explanation}(c). 
%
This behavior can be attributed to the energetic alignment between the molecular frontier orbitals and the surface Fermi level in the CPLX system.
In particular, $\epsilon_{\rm H,\rm DM}$ plays a critical role, as evidenced by its stronger correlation with $E_{\rm f,CPLX}$ (see Fig. \ref{explanation}(b)). We therefore hypothesize that when $\epsilon_{\rm H,\rm DM}$ lies well below the surface Fermi level, the HOMO is energetically inaccessible and induces negligible charge redistribution at the surface, resulting in a minimal impact on work-function modulation.
Conversely, when $\epsilon_{\rm H,\rm DM}$ exceeds the surface Fermi level, substantial charge redistribution can occur, and $\Delta\phi$ is governed by the energetic separation between the HOMO and the surface Fermi level.
Finally, this threshold effect may also be influenced by the spatial localization of the frontier orbitals on the molecule \cite{egger2013anticorrelation_dwf_homo}, which affects their coupling to the surface. A detailed analysis of these spatial effects, however, is beyond the scope of the present work.

\section*{Discussion}

In the present work, we introduce MORE-ML, a computational framework that integrates quantum-mechanical (QM) property data of electronic-nose molecular building blocks with machine-learning (ML) methods to predict and interpret the physicochemical mechanisms governing sensing-related properties.
This challenging task is addressed by expanding our previously generated MORE-Q dataset into MORE-QX, which spans a significantly larger conformational and property space for interacting systems composed of combinations of body-odor volatilomes (BOVs) and mucin-derived receptors (REC).
Based on MORE-QX, we construct a set of binding features (BFs) by computing the adsorption energy ($E_{\rm ads}$), work-function change ($\Delta \phi$), and charge transfer ($\Delta Q$). These quantities quantify the impact of BOV–REC interactions on the energy, work function ($\phi$), and charge distribution of the REC–graphene systems.
Analysis of the property space spanned by MORE-QX reveals clear evidence of ``Freedom of design'' in the BF space, \ie the ability to identify chemically diverse OM–REC–graphene (CPLX) conformations that exhibit a targeted set of BFs. This flexibility arises from the weak correlations observed among most QM properties.
Furthermore, property–property correlation analysis highlights the potential of several electronic features to discriminate between similar DM and CPLX conformations, a key requirement for constructing efficient molecular descriptors.
Most electronic features included in MORE-QX are invariant with respect to translations, rotations, and atom permutations, thereby satisfying a central requirement for a complete molecular representation suitable for ML-based predictive modeling.

Leveraging these insights within the MORE-ML framework, we define deterministic mappings between the electronic features of molecular building blocks (\eg OM, REC, and DM systems) and the BFs. 
These mappings are designed to reduce the computational cost of determining sensing-related properties, as computing QM properties for individual building blocks is significantly less expensive than direct BF calculations.
To this end, we performed feature engineering and benchmark multiple ML regression techniques to identify the optimal set of electronic features for developing accurate and reliable regression models for each BF.
In contrast to previous ML studies that primarily emphasize predictive performance, we place strong emphasis on model explainability by combining the intrinsic interpretability of tree-based models with SHapley Additive exPlanations (SHAP) analysis.
Indeed, we find that $E_{\rm ads}$ is largely governed by the interaction energy between the OM and REC systems, whereas $\Delta Q$ is primarily influenced by charge-related properties, such as dipole and quadrupole moments. In the case of $\Delta \phi$, an interplay emerges between the vertical dipole moment of OM and the HOMO energy of the DM system, reflecting the physical mechanisms underlying the determination of the work function $\phi$.
This in-depth investigation reveals the key physicochemical factors governing each BF and thereby establishes a more transparent and navigable pathway through the largely unexplored binding feature space.

From the electronic-nose sensing materials design perspective, the demonstrated ``Freedom of design'' in the binding feature space is particularly valuable, as it suggests that sensor sensitivity, baseline stability, and selectivity can be tuned semi-independently through receptor engineering rather than relying on trial-and-error material screening. The finding that adsorption energy, charge transfer, and work function modulation are governed by distinct and weakly correlated electronic descriptors aligns well with practical observations in sensor arrays, where signal amplitude, recovery behavior, and device-to-device variability often decouple. Importantly, the interpretability of the MORE-ML framework provides experimentally actionable guidelines for selecting or synthesizing receptor molecules that target specific transduction mechanisms, thereby reducing empirical optimization cycles. This QM-ML-experiment feedback loop represents a critical step toward rational, scalable design of next-generation digital olfaction systems.

Based on the findings presented in this work, we successfully demonstrate a sustainable AI-based framework that reveals multiple sensing mechanisms from a computational perspective.
Although MORE-QX is limited to sensing-related properties on graphene surfaces, this comprehensive analysis elucidates the fundamental mechanisms controlling BFs---properties that are strongly linked to sensing performance---through the manipulation of DM dimer properties.
These insights pave the way for defining novel design principles for high-performance, sensitive, and selective molecular receptors, which can be validated using generative AI approaches or experimental measurements.
Moreover, the understanding gained in this work can be directly transferred to more practically relevant sensing materials, such as two-dimensional MXenes \cite{cai2025recent_mxene_sensing,yu20252d_mxene_2}, transition-metal dichalcogenides (TMDs) \cite{mirzaei2024resistive_TMDs_mxene,Jana2025} or metal-organic frameworks (MOFs)\cite{Wang2025}, which offer a richer chemical space and enhanced electronic tunability for gas-sensing applications.
We note that achieving a full understanding of sensing mechanisms in electronic-nose devices also requires investigating the contact potential between the electrode and the sensing surface (\ie the Schottky barrier effect \cite{mathew2021schottky}), as it may play a dominant role in sensing performance.
Therefore, we expect this work to motivate future research aimed at advancing sensing materials by leveraging physical and chemical insights together with deterministic property mappings enabled by the integration of quantum science and interpretable ML regression models.

\section*{Methods}
\subsection*{DFT computational details}
The QM properties of BOVs, molecular receptors, and their dimer conformations were obtained both at GFN2-xTB+D4 level and PBE+D3 with def2-TZVPP basis set using xTB (version 6.6.0) \cite{bannwarth2019xtb} and ORCA (version 5.0.3)\cite{neese2012orca} packages, respectively.
In MORE-QX, the dimer interaction energy $E_{\rm int}$ is defined as the total energy of the dimer conformation minus the energies of the individual constituents, \ie 

\begin{equation}
    E_{\rm int} = E_{\rm DM} - E_{\rm REC} - E_{\rm OM}.
\end{equation}

The BOV-receptor-graphene complex (CPLX) systems underwent geometry optimization using the DFTB+\cite{hourahine2020dftb} package, employing the GFN2-xTB Hamiltonian with D4 dispersion correction. 
While optimizing the structures, we fixed the atomic positions in the graphene layer, as the adsorption of the OM molecules will not significantly affect the geometry of graphene, and the electrode is restricting the deformation degree of the graphene for a chemiresistive sensing device. 
To create the SUB system (or REC-graphene system), we removed the BOV molecule from the CPLX system and did not optimize the structures in order to investigate the pure electronic effect of the binding features. 

Similar to the MORE-Q dataset\cite{more-q}, MORE-QX provides extensive sets of QM global and local properties (up to 39) for single BOV/receptor molecules (MORE-QX-G1), BOV-receptor molecular dimers (MORE-QX-G2), and complex systems (MORE-QX-G3). 
The MORE-QX-G1 subset contains QM property data for 102 BOV molecules and 18 molecular receptors. Among the 39 molecular and atomic properties, we computed the D3 energy, dipole moment, polarizability, and Mulliken charges. The MORE-QX-G2 subset is built on the geometries from MORE-QX-G1 via the search for molecular docking conformations using BOV molecules and receptors. Accordingly, MORE-QX-G2 contains QM property data for 23,838 dimer conformations at the GFN2-xTB+D4 level and for 10,411 dimers with the lowest binding energies at the PBE+D3 level (see the property list in Tables S2 and S3 of the SI). The MORE-QX-G3 subset contains 10,411 selected dimers from MORE-QX-G2 on graphene surface. Consequently, MORE-Q-G3 includes QM property data at the PBE+D3 level for both the CPLX and SUB systems, as well as binding features that account for property changes in single systems induced by BOV molecule adsorption.
%
The expansion including GFN-xTB+D4 geometry relaxation and DFT calculation took $\sim 25$ Mio CPUhs.

To measure the correlation between QM properties in MORE-QX, we have used the Spearman correlation factor, which is computed as follow: 
\begin{equation}
    |\rho_s| = |1 - \frac{6\sum^n_{i=1}d^2_i}{n(n^2-1)}|,
    \label{spearmann}
\end{equation}
where each paired observation $(X_i,Y_i)$'s respective ranks denotes $R(X_i)$ and $R(Y_i)$, and then the $d_i$ is defined as $d_i = R(X_i) - R(Y_i)$.
Spearman is chosen owing to its robustness against outliers and the enhanced non-linear capturing ability compared to the counterpart Pearson correlation.  

\subsection*{Binding feature calculation}
Electronic-structure calculations of the SUB and CPLX systems were conducted at tightly converged
PBE+D3 theory level by Vienna ab initio simulation package (VASP\cite{kresse1996vasp_1,kresse1996vasp_2}, version 6.3.1). The energy cutoff for the plane-wave basis set and the SCF convergence threshold were set to $600$ and $1\cdot10^{-4}$ eV, respectively. And all simulations were conducted at Gamma point. The dipole correction along the slab direction (50.68\AA) was switched on to obtain flat electrostatic potential. 

To compute the binding features, we carried out different type of DFT calculations. The adsorption energy ($E_{\mathrm{ads}}$) was obtained from total energies of single-point calculations and is defined as follows:
\begin{equation}
    E_{\mathrm{ads}} = E_{\mathrm{CPLX}} - E_{\mathrm{SUB}} - E_{\mathrm{OM}}.
\end{equation}

Whereas, the work function change ($\Delta \phi$) is defined as the difference between the work function ($\phi$)  after and before the BOV adsorption, \ie $\phi$ for CPLX and SUB systems:
\begin{equation}
    \Delta \phi = \phi_{\mathrm{CPLX}} - \phi_{\mathrm{SUB}}.
    \label{Work function change}
\end{equation}
Here, $\phi$ of each system was calculated using:
\begin{equation}
    \phi = E_{\mathrm{V}} - E_{\mathrm{F}},
    \label{work function equation}
\end{equation}
where $E_{\rm F}$ is the Fermi level and $E_{\rm V}$ is the vacuum energy. $E_{\rm V}$ is obtained by analyzing the flattened region of the electrostatic potential $P(z)$ along the slab direction.  
$P(z)$ is computed by the following equation: 
\begin{equation}
P(z) = \int n(z) dz,
\label{ELC}
\end{equation}
where the planar averaged charge density $n(z)$ is defined as:
\begin{equation}
n(z) = 1/A \iint n(x,y,z) dxdy 
\label{PLA}
\end{equation}
Finally, the charge transfer $\Delta Q$ is computed as the total Bader charge \cite{henkelman2006bader} transferring between the BOV molecule and SUB system.

\subsection*{Conformer sampling}
The initial 83,916 dimer configurations (50 configurations per combination) were searched by automated Interaction Site Screening (aISS) package \cite{plett2023automated_aiss}.
Then we conducted the geometrical root-mean-squared-deviation (RMSD)-based hierarchical clustering, where we set the cut-off RMSD distance to filter the geometrically redundant configurations on the whole 83,916 configurations level, which led to 23,838 configurations. 
As a result, simple-geometry binding configurations are scarce, whereas complex-geometry configurations are abundant in the remaining 23,838 dimer configurations.
Therefore, when depositing low-energy conformers onto graphene surface (evaluated by the interaction energy $E_{\rm int}$) in this work, complex conformers are sampled more frequently than simple ones, resulting in an average of six conformers per dimer combination. 
More computational details can be found in Ref. \cite{more-q}.

\subsection*{MORE-ML framework}

We designed the \textbf{M}olecular \textbf{O}lfactorial \textbf{R}eceptor \textbf{E}ngineering by \textbf{M}achine \textbf{L}earning (\textbf{MORE-ML}) framework to simultaneously perform binding feature regression and model explanation tasks, as illustrated in Fig. \ref{ML_workflow}(a).
Among the spectrum of ML algorithms, linear models offer the highest explainability but lack sufficient capacity, whereas neural networks provide exceptional representational power yet suffer from nascent explainability \cite{adadi2018explain_general}.
To strike a balance between predictive performance and transparency, we employ tree-based models, which deliver both robust accuracy and an inherently interpretable decision process via hierarchical splitting \cite{lundberg2020_tree_model_explaination}. 
Moreover, when integrated with explainable artificial intelligence (XAI) tools--\eg \textbf{SH}apley \textbf{A}dditive ex\textbf{P}lanations (SHAP) \cite{SHAP2017}--these models not only yield precise predictions of binding features but also facilitate the extraction of underlying physical insights \cite{oviedo2022interpretable_materials_intro}.

Building on the defined ML tasks, we now describe the training procedure for a single ML model, as depicted in Fig.~\ref{ML_workflow}(b).
In our initial training loops, we identified some systems in which the dominant interactions occurred between the OM and the graphene surface rather than with the receptor. 
By projecting the data into UMAP space and clustering based on SHAP values (for better cluster forming \cite{cooper2021supervised_shap_value_cluster,usuga2024local_shap_cluster}), we uncovered a distinct cluster corresponding to these outliers. We subsequently removed all 932 systems, as they lie outside the scope of DM pair design and would otherwise impair the effectiveness of our model (see more details in Fig. S4 of the SI).

The remaining data points are then split into training and test sets via farthest‐point sampling (FPS) in the binding‐feature t-SNE space, since t-SNE captures nonlinear relationships and clusters systems with similar binding mechanisms—preserving local consistency better than alternatives such as PCA or UMAP and homogeneous sampling in this space minimizes distributional divergence between the two sets.
The dataset was partitioned into training and test sets at a fixed ratio of $9\!:\!1$. The corresponding learning curves are shown in Fig.  S5.
This fixed test set is used to benchmark both intermediate models and the final model throughout the entire training process. 
Then we conducted 100 iterations of Bayesian optimization (BO) to identify optimal hyperparameters, using the mean root‐mean‐square error (mRMSE) from 10‐fold cross‐validation at each BO iteration as the objective. The best‐found hyperparameters were then applied to retrain the models on the training set, and final performance was evaluated on the fixed test set.
%

\subsection*{Explainability strategy for tree-based regression models}
In this work, we employ \textbf{SH}apley \textbf{A}dditive ex\textbf{P}lanations (SHAP) to interpret ML regression models developed for predicting binding features.
SHAP is a game-theoretic framework for explaining ML model outputs, grounded in cooperative game theory and based on Shapley values, which quantify how each input feature influences the deviation of an individual prediction from the expected/average output of the model. 
Consequently, this method allows for a more transparent interpretation of the learned correlations, highlighting the relative importance of features and how they interact to affect the predicted outcomes. 
%
SHAP converts the value of feature $j$ to the SHAP value $\phi_j$ by considering its margin contribution towards the model $f$ output, and hence the SHAP value of feature $j$ is defined as:
\begin{equation}
    \phi_j(f) = \sum_{S\subseteq N \textbackslash \{j\}} \frac{|S|!(|N|-|S|-1)!}{|N|!} [f(S \cup \{j\}) - f\{S\}],
\end{equation}
where $S$ stands for feature subset without feature $j$, $N$ is the total feature set, and $f$ is the ML model. 
This equation defines the SHAP value as the sum of feature $j$'s marginal contributions across every subset $S$, each term weighted by the probability that exactly those features in $S$ appear before $j$ in all ordering combinations of all features.

In the same context, we also use the intrinsic explainability of decision-tree–based models, which formulate predictions as a nested rule structure. Starting from the root node, the model recursively subdivides the feature space by applying feature-dependent thresholding conditions (\eg border values in CatBoost), producing a hierarchy of progressively constrained decision subspaces. The partitioning process terminates at leaf nodes, each associated with a fixed prediction value or a set of distributional parameters. The prediction mechanism for any leaf can be explicitly recovered by back-tracking along its unique partition path, yielding an interpretable representation of the model as a piecewise-constant function over disjoint regions of the input space.

\bibliography{sample}

@article{more-q,
  title={{MORE-Q}, a dataset for molecular olfactorial receptor engineering by quantum mechanics},
  author={Chen, Li and Medrano Sandonas, Leonardo and Traber, Philipp and Dianat, Arezoo and Tverdokhleb, Nina and Hurevich, Mattan and Yitzchaik, Shlomo and Gutierrez, Rafael and Croy, Alexander and Cuniberti, Gianaurelio},
  journal={Scientific Data},
  volume={12},
  number={1},
  pages={324},
  year={2025},
  publisher={Nature Publishing Group UK London}
}

@article{DFT_ads_process_2011,
  title={Density functional theory in surface chemistry and catalysis},
  author={N{\o}rskov, Jens K and Abild-Pedersen, Frank and Studt, Felix and Bligaard, Thomas},
  journal={Proceedings of the National Academy of Sciences},
  volume={108},
  number={3},
  pages={937--943},
  year={2011},
  publisher={National Academy of Sciences}
}

@article{DWF_DIP_Li,
  title={Computational Design of the Electronic Response for Volatile Organic Compounds Interacting with Doped Graphene Substrates},
  author={Chen, Li and Bodesheim, David and Ranjbar, Ahmad and Dianat, Arezoo and Biele, Robert and Gutierrez, Rafael and Khazaei, Mohammad and Cuniberti, Gianaurelio},
  journal={Nanomaterials},
  volume={14},
  number={22},
  pages={1778},
  year={2024},
  publisher={MDPI}
}

@article{DWF_mohannmad,
  title={{OH}-terminated two-dimensional transition metal carbides and nitrides as ultralow work function materials},
  author={Khazaei, Mohammad and Arai, Masao and Sasaki, Taizo and Ranjbar, Ahmad and Liang, Yunye and Yunoki, Seiji},
  journal={Physical Review B},
  volume={92},
  number={7},
  pages={075411},
  year={2015},
  publisher={APS}
}

@article{DWF_leung,
  title={Relationship between surface dipole, work function and charge transfer: Some exceptions to an established rule},
  author={Leung, Tsan-Chuen and Kao, CL and Su, WS and Feng, YJ and Chan, Che Ting},
  journal={Physical Review B},
  volume={68},
  number={19},
  pages={195408},
  year={2003},
  publisher={APS}
}

@article{2015_orbital_mixing,
  title={Mechanism of charge transfer and its impacts on Fermi-level pinning for gas molecules adsorbed on monolayer {WS}$_2$},
  author={Zhou, Changjie and Yang, Weihuang and Zhu, Huili},
  journal={The Journal of chemical physics},
  volume={142},
  number={21},
  year={2015},
  publisher={AIP Publishing}
}

@incollection{SHAP2017,
title = {A Unified Approach to Interpreting Model Predictions},
author = {Lundberg, Scott M and Lee, Su-In},
booktitle = {Advances in Neural Information Processing Systems 30},
editor = {I. Guyon and U. V. Luxburg and S. Bengio and H. Wallach and R. Fergus and S. Vishwanathan and R. Garnett},
pages = {4765--4774},
year = {2017},
publisher = {Curran Associates, Inc.},
url = {http://papers.nips.cc/paper/7062-a-unified-approach-to-interpreting-model-predictions.pdf}
}

@article{zhang2020emotion,
  title={Emotion recognition using multi-modal data and machine learning techniques: {A} tutorial and review},
  author={Zhang, Jianhua and Yin, Zhong and Chen, Peng and Nichele, Stefano},
  journal={Inf. Fusion},
  volume={59},
  pages={103--126},
  year={2020},
  publisher={Elsevier}
}

@article{ali2020_enose,
  title={Principles and recent advances in electronic nose for quality inspection of agricultural and food products},
  author={Ali, Maimunah Mohd and Hashim, Norhashila and Abd Aziz, Samsuzana and Lasekan, Ola},
  journal={Trends Food Sci. Technol.},
  volume={99},
  pages={1--10},
  year={2020},
  publisher={Elsevier}
}

@article{drabinska_bovs,
  title={A literature survey of all volatiles from healthy human breath and bodily fluids: {T}he human volatilome},
  author={Drabi{\'n}ska, Natalia and Flynn, Cheryl and Ratcliffe, Norman and Belluomo, Ilaria and Myridakis, Antonis and Gould, Oliver and Fois, Matteo and Smart, Amy and Devine, Terry and Costello, Ben De Lacy},
  journal={Journal of Breath Research},
  volume={15},
  number={3},
  pages={034001},
  year={2021},
  publisher={IOP Publishing}
}

@article{tisch2013detection,
  title={Detection of {A}lzheimer's and {P}arkinson's disease from exhaled breath using nanomaterial-based sensors},
  author={Tisch, Ulrike and Schlesinger, Ilana and Ionescu, Radu and Nassar, Maria and Axelrod, Noa and Robertman, Dorina and Tessler, Yael and Azar, Faris and Marmur, Abraham and Aharon-Peretz, Judith and others},
  journal={Nanomedicine},
  volume={8},
  number={1},
  pages={43--56},
  year={2013},
  publisher={Taylor \& Francis}
}

@article{trivedi2019_parkinson,
  title={Discovery of volatile biomarkers of {P}arkinson’s disease from sebum},
  author={Trivedi, Drupad K and Sinclair, Eleanor and Xu, Yun and Sarkar, Depanjan and Walton-Doyle, Caitlin and Liscio, Camilla and Banks, Phine and Milne, Joy and Silverdale, Monty and Kunath, Tilo and others},
  journal={ACS Cent. Sci.},
  volume={5},
  number={4},
  pages={599--606},
  year={2019},
  publisher={ACS Publications}
}

@article{2016_sp_smeller,
  title={Joy of super smeller: sebum clues for {PD} diagnostics},
  author={Morgan, Jules},
  journal={Lancet Neurol.},
  volume={15},
  number={2},
  pages={138--139},
  year={2016},
  publisher={Elsevier}
}

@article{2001_OR,
  title={How the olfactory system makes sense of scents},
  author={Firestein, Stuart},
  journal={Nature},
  volume={413},
  number={6852},
  pages={211--218},
  year={2001},
  publisher={Nature Publishing Group UK London}
}

@article{bakhatan2023_rec1,
  title={Accelerated solid phase glycan synthesis: ASGS},
  author={Bakhatan, Yasmeen and Alshanski, Israel and Chan, Chieh-Kai and Lo, Wei-Chih and Lu, Po-Wei and Liao, Pin-Hsuan and Wang, Cheng-Chung and Hurevich, Mattan},
  journal={Chemistry--A European Journal},
  volume={29},
  number={38},
  pages={e202300897},
  year={2023},
  publisher={Wiley Online Library}
}

@article{sukhran2023_rec2,
  title={Unexpected Nucleophile Masking in Acyl Transfer to Sterically Crowded and Conformationally Restricted Galactosides},
  author={Sukhran, Yonatan and Alshanski, Israel and Filiba, Ofer and Mackintosh, Megan J and Schapiro, Igor and Hurevich, Mattan},
  journal={J. Org. Chem.},
  volume={88},
  number={13},
  pages={9313--9320},
  year={2023},
  publisher={ACS Publications}
}

@article{wehling2008_CT,
  title={Molecular doping of graphene},
  author={Wehling, TO and Novoselov, KS and Morozov, SV and Vdovin, EE and Katsnelson, MI and Geim, AK and Lichtenstein, AI},
  journal={Nano letters},
  volume={8},
  number={1},
  pages={173--177},
  year={2008},
  publisher={ACS Publications}
}

@article{mathew2021schottky,
  title={Schottky diodes based on 2{D} materials for environmental gas monitoring: a review on emerging trends, recent developments and future perspectives},
  author={Mathew, Minu and Rout, Chandra Sekhar},
  journal={Journal of Materials Chemistry C},
  volume={9},
  number={2},
  pages={395--416},
  year={2021},
  publisher={Royal Society of Chemistry}
}

@article{zhang2009_recovery_time,
  title={Improving gas sensing properties of graphene by introducing dopants and defects: {A} first-principles study},
  author={Zhang, Yong-Hui and Chen, Ya-Bin and Zhou, Kai-Ge and Liu, Cai-Hong and Zeng, Jing and Zhang, Hao-Li and Peng, Yong},
  journal={Nanotechnology},
  volume={20},
  number={18},
  pages={185504},
  year={2009},
  publisher={IOP Publishing}
}

@article{jeindl2022polymorphism,
  title={How much does surface polymorphism influence the work function of organic/metal interfaces?},
  author={Jeindl, Andreas and H{\"o}rmann, Lukas and Hofmann, Oliver T},
  journal={Applied Surface Science},
  volume={575},
  pages={151687},
  year={2022},
  publisher={Elsevier}
}

@article{chen2025multi_ads1,
  title={A multi-modal transformer for predicting global minimum adsorption energy},
  author={Chen, Junwu and Huang, Xu and Hua, Cheng and He, Yulian and Schwaller, Philippe},
  journal={Nature Communications},
  volume={16},
  number={1},
  pages={3232},
  year={2025},
  publisher={Nature Publishing Group UK London}
}

@article{lan2023adsorbml_ads1,
  title={Adsorb{ML}: a leap in efficiency for adsorption energy calculations using generalizable machine learning potentials},
  author={Lan, Janice and Palizhati, Aini and Shuaibi, Muhammed and Wood, Brandon M and Wander, Brook and Das, Abhishek and Uyttendaele, Matt and Zitnick, C Lawrence and Ulissi, Zachary W},
  journal={npj Computational Materials},
  volume={9},
  number={1},
  pages={172},
  year={2023},
  publisher={Nature Publishing Group UK London}
}

@article{pablo2023fast-gamenet,
  title={Fast evaluation of the adsorption energy of organic molecules on metals via graph neural networks},
  author={Pablo-Garc{\'\i}a, Sergio and Morandi, Santiago and Vargas-Hern{\'a}ndez, Rodrigo A and Jorner, Kjell and Ivkovi{\'c}, {\v{Z}}arko and L{\'o}pez, N{\'u}ria and Aspuru-Guzik, Al{\'a}n},
  journal={Nature Computational Science},
  volume={3},
  number={5},
  pages={433--442},
  year={2023},
  publisher={Nature Publishing Group US New York}
}

@article{tran2018_ads_10,
  title={Active learning across intermetallics to guide discovery of electrocatalysts for \ce{CO2} reduction and \ce{H2} evolution},
  author={Tran, Kevin and Ulissi, Zachary W},
  journal={Nature Catalysis},
  volume={1},
  number={9},
  pages={696--703},
  year={2018},
  publisher={Nature Publishing Group UK London}
}

@article{zhong2020_ads11,
  title={Accelerated discovery of \ce{CO2} electrocatalysts using active machine learning},
  author={Zhong, Miao and Tran, Kevin and Min, Yimeng and Wang, Chuanhao and Wang, Ziyun and Dinh, Cao-Thang and De Luna, Phil and Yu, Zongqian and Rasouli, Armin Sedighian and Brodersen, Peter and others},
  journal={Nature},
  volume={581},
  number={7807},
  pages={178--183},
  year={2020},
  publisher={Nature Publishing Group UK London}
}

@article{fung2021_ads12,
  title={Machine learned features from density of states for accurate adsorption energy prediction},
  author={Fung, Victor and Hu, Guoxiang and Ganesh, Panchapakesan and Sumpter, Bobby G},
  journal={Nature Commun.},
  volume={12},
  number={1},
  pages={88},
  year={2021},
  publisher={Nature Publishing Group UK London}
}

@article{xu2022predicting_ads13,
  title={Predicting binding motifs of complex adsorbates using machine learning with a physics-inspired graph representation},
  author={Xu, Wenbin and Reuter, Karsten and Andersen, Mie},
  journal={Nature Computational Science},
  volume={2},
  number={7},
  pages={443--450},
  year={2022},
  publisher={Nature Publishing Group US New York}
}

@article{li2024interpreting_ads14,
  title={Interpreting chemisorption strength with AutoML-based feature deletion experiments},
  author={Li, Zhuo and Zhao, Changquan and Wang, Haikun and Ding, Yanqing and Chen, Yechao and Schwaller, Philippe and Yang, Ke and Hua, Cheng and He, Yulian},
  journal={Proceedings of the National Academy of Sciences},
  volume={121},
  number={12},
  pages={e2320232121},
  year={2024},
  publisher={National Academy of Sciences}
}

@article{bhati2021gas_sensing_mechanism,
  title={Gas sensing performance of 2D nanomaterials/metal oxide nanocomposites: A review},
  author={Bhati, Vijendra Singh and Kumar, Mahesh and Banerjee, Rupak},
  journal={Journal of Materials Chemistry C},
  volume={9},
  number={28},
  pages={8776--8808},
  year={2021},
  publisher={Royal Society of Chemistry}
}

@article{sandonas2023freedom,
  title={“{Freedom of design}” in chemical compound space: towards rational in silico design of molecules with targeted quantum-mechanical properties},
  author={Medrano Sandonas, Leonardo and Hoja, Johannes and Ernst, Brian G and V{\'a}zquez-Mayagoitia, {\'A}lvaro and DiStasio, Robert A and Tkatchenko, Alexandre},
  journal={Chemical Science},
  volume={14},
  number={39},
  pages={10702--10717},
  year={2023},
  publisher={Royal Society of Chemistry}
}

@book{kreuzer2012physisorption,
  title={Physisorption kinetics},
  author={Kreuzer, Hans J{\"u}rgen and Gortel, Zbigniew W},
  volume={1},
  year={2012},
  publisher={Springer Science \& Business Media}
}

@article{kim2025functional,
  title={Functional monomer design for synthetically accessible polymers},
  author={Kim, Seonghwan and Schroeder, Charles M and Jackson, Nicholas E},
  journal={Chemical Science},
  volume={16},
  number={11},
  pages={4755--4767},
  year={2025},
  publisher={Royal Society of Chemistry}
}

@article{lundberg2020_tree_model_explaination,
  title={From local explanations to global understanding with explainable {AI} for trees},
  author={Lundberg, Scott M and Erion, Gabriel and Chen, Hugh and DeGrave, Alex and Prutkin, Jordan M and Nair, Bala and Katz, Ronit and Himmelfarb, Jonathan and Bansal, Nisha and Lee, Su-In},
  journal={Nat. Mach. Intell.},
  volume={2},
  number={1},
  pages={56--67},
  year={2020},
  publisher={Nature Publishing Group}
}

@article{adadi2018explain_general,
  title={Peeking inside the black-box: a survey on explainable artificial intelligence ({XAI})},
  author={Adadi, Amina and Berrada, Mohammed},
  journal={IEEE access},
  volume={6},
  pages={52138--52160},
  year={2018},
  publisher={IEEE}
}

@article{oviedo2022interpretable_materials_intro,
  title={Interpretable and explainable machine learning for materials science and chemistry},
  author={Oviedo, Felipe and Ferres, Juan Lavista and Buonassisi, Tonio and Butler, Keith T},
  journal={Accounts of Materials Research},
  volume={3},
  number={6},
  pages={597--607},
  year={2022},
  publisher={ACS Publications}
}

@article{bannwarth2019xtb,
  title={{GFN2-xTB—An} accurate and broadly parametrized self-consistent tight-binding quantum chemical method with multipole electrostatics and density-dependent dispersion contributions},
  author={Bannwarth, Christoph and Ehlert, Sebastian and Grimme, Stefan},
  journal={Journal of chemical theory and computation},
  volume={15},
  number={3},
  pages={1652--1671},
  year={2019},
  publisher={ACS Publications}
}

@article{henkelman2006bader,
  title={A fast and robust algorithm for Bader decomposition of charge density},
  author={Henkelman, Graeme and Arnaldsson, Andri and J{\'o}nsson, Hannes},
  journal={Computational Materials Science},
  volume={36},
  number={3},
  pages={354--360},
  year={2006},
  publisher={Elsevier}
}

@article{neese2012orca,
  title={{The ORCA} program system},
  author={Neese, Frank},
  journal={Wiley Interdisciplinary Reviews: Computational Molecular Science},
  volume={2},
  number={1},
  pages={73--78},
  year={2012},
  publisher={Wiley Online Library}
}

@article{kresse1996vasp_1,
  title={Efficient iterative schemes for ab initio total-energy calculations using a plane-wave basis set},
  author={Kresse, Georg and Furthm{\"u}ller, J{\"u}rgen},
  journal={Phys. Rev. B},
  volume={54},
  number={16},
  pages={11169},
  year={1996},
  publisher={APS}
}

@article{kresse1996vasp_2,
  title={Efficiency of ab-initio total energy calculations for metals and semiconductors using a plane-wave basis set},
  author={Kresse, Georg and Furthm{\"u}ller, J{\"u}rgen},
  journal={Comput. Mater. Sci.},
  volume={6},
  number={1},
  pages={15--50},
  year={1996},
  publisher={Elsevier}
}

@article{kovacs2025mace_off,
  title={Mace-off: Short-range transferable machine learning force fields for organic molecules},
  author={Kov{\'a}cs, D{\'a}vid P{\'e}ter and Moore, J Harry and Browning, Nicholas J and Batatia, Ilyes and Horton, Joshua T and Pu, Yixuan and Kapil, Venkat and Witt, William C and Magdau, Ioan-Bogdan and Cole, Daniel J and others},
  journal={Journal of the American Chemical Society},
  volume={147},
  number={21},
  pages={17598--17611},
  year={2025},
  publisher={ACS Publications}
}

@article{breiman2001RF,
  title={Random forests},
  author={Breiman, Leo},
  journal={Machine learning},
  volume={45},
  number={1},
  pages={5--32},
  year={2001},
  publisher={Springer}
}

@article{ke2017lightgbm,
  title={Lightgbm: A highly efficient gradient boosting decision tree},
  author={Ke, Guolin and Meng, Qi and Finley, Thomas and Wang, Taifeng and Chen, Wei and Ma, Weidong and Ye, Qiwei and Liu, Tie-Yan},
  journal={Adv. Neural Inf. Process. Syst.},
  volume={30},
  year={2017}
}

@inproceedings{chen2016xgboost,
  title={Xgboost: A scalable tree boosting system},
  author={Chen, Tianqi and Guestrin, Carlos},
  booktitle={Proceedings of the 22nd acm sigkdd international conference on knowledge discovery and data mining},
  pages={785--794},
  year={2016}
}

@article{prokhorenkova2018catboost,
  title={CatBoost: unbiased boosting with categorical features},
  author={Prokhorenkova, Liudmila and Gusev, Gleb and Vorobev, Aleksandr and Dorogush, Anna Veronika and Gulin, Andrey},
  journal={Adv. Neural Inf. Process. Syst.},
  volume={31},
  year={2018}
}

@article{cuniberti2005_molecular_electronics,
  title={Introducing molecular electronics: A brief overview},
  author={Cuniberti, Gianaurelio and Fagas, Giorgos and Richter, Klaus},
  journal={Introducing molecular electronics},
  pages={1--10},
  year={2005},
  publisher={Springer}
}

@article{egger2013anticorrelation_dwf_homo,
  title={Anticorrelation between the evolution of molecular dipole moments and induced work function modifications},
  author={Egger, David A and Zojer, Egbert},
  journal={J. Phys. Chem. Lett.},
  volume={4},
  number={20},
  pages={3521--3526},
  year={2013},
  publisher={ACS Publications}
}

@article{plett2023automated_aiss,
  title={Automated and efficient generation of general molecular aggregate structures},
  author={Plett, Christoph and Grimme, Stefan},
  journal={Angew. Chem. Int. Ed.},
  volume={62},
  number={4},
  pages={e202214477},
  year={2023},
  publisher={Wiley Online Library}
}

@inproceedings{cooper2021supervised_shap_value_cluster,
  title={Supervised clustering for subgroup discovery: an application to {COVID}-19 symptomatology},
  author={Cooper, Aidan and Doyle, Orla and Bourke, Alison},
  booktitle={Joint European conference on machine learning and knowledge discovery in databases},
  pages={408--422},
  year={2021},
  organization={Springer}
}

@article{hinostroza25, title={Assessing the performance of quantum-mechanical descriptors in physicochemical and biological property prediction}, DOI={10.26434/chemrxiv-2025-hj4dc}, journal={ChemRxiv}, author={Hinostroza Caldas, Alejandra and Kokorin, Artem and Tkatchenko, Alexandre and Medrano Sandonas, Leonardo}, year={2025}}

@article{puleva2025,
	author = {Puleva, Mirela and Medrano Sandonas, Leonardo and L{\H o}rincz, Bal{\'a}zs D. and Charry, Jorge and Rogers, David M. and Nagy, P{\'e}ter R. and Tkatchenko, Alexandre},
	journal = {Nature Communications},
	number = {1},
	pages = {8583},
	title = {Extending quantum-mechanical benchmark accuracy to biological ligand-pocket interactions},
	volume = {16},
	year = {2025}}

@article{aqm,
	author = {Medrano Sandonas, Leonardo and Van Rompaey, Dries and Fallani, Alessio and Hilfiker, Mathias and Hahn, David and Perez-Benito, Laura and Verhoeven, Jonas and Tresadern, Gary and Kurt Wegner, Joerg and Ceulemans, Hugo and Tkatchenko, Alexandre},
	journal = {Scientific Data},
	number = {1},
	pages = {742},
	title = {Dataset for quantum-mechanical exploration of conformers and solvent effects in large drug-like molecules},
	volume = {11},
	year = {2024}}

@article{usuga2024local_shap_cluster,
  title={Local descriptors-based machine learning model refined by cluster analysis for accurately predicting adsorption energies on bimetallic alloys},
  author={Usuga, AF and Praveen, CS and Comas-Vives, A},
  journal={Journal of Materials Chemistry A},
  volume={12},
  number={5},
  pages={2708--2721},
  year={2024},
  publisher={Royal Society of Chemistry}
}

@article{cai2025recent_mxene_sensing,
  title={Recent advances in MXene gas sensors: synthesis, composites, and mechanisms},
  author={Cai, Zhicheng and Kim, Hyojung},
  journal={npj 2D Materials and Applications},
  volume={9},
  number={1},
  pages={66},
  year={2025},
  publisher={Nature Publishing Group UK London}
}

@article{yu20252d_mxene_2,
  title={{2D MXenes-Based Gas Sensors: Progress, Applications, and Challenges}},
  author={Yu, Shuguo and Li, Peng and Ding, Hanqin and Liang, Chongyu and Wang, Xiaojun},
  journal={Small Methods},
  pages={2402179},
  year={2025},
  publisher={Wiley Online Library}
}

@article{mirzaei2024resistive_TMDs_mxene,
  title={{Resistive gas sensors based on 2D TMDs and MXenes}},
  author={Mirzaei, Ali and Kim, Jin-Young and Kim, Hyoun Woo and Kim, Sang Sub},
  journal={Accounts of Chemical Research},
  volume={57},
  number={16},
  pages={2395--2413},
  year={2024},
  publisher={ACS Publications}
}

@article{shitrit2025monosaccharide,
  title={Monosaccharide-Derived Enantioselectivity in SWCNT Chemoresistive VOC Sensing},
  author={Shitrit, Ariel and Sukhran, Yonatan and Tverdokhleb, Nina and Chen, Li and Dianat, Arezoo and Gutierrez, Rafael and K{\"o}rbel, Sabine and Croy, Alexander and Cuniberti, Gianaurelio and Hurevich, Mattan and others},
  journal={Chemistry--A European Journal},
  pages={e02553},
  year={2025},
  publisher={Wiley Online Library}
}

@article{huang2022_enose,
  title={Machine learning-enabled smart gas sensing platform for identification of industrial gases},
  author={Huang, Shirong and Croy, Alexander and Panes-Ruiz, Luis Antonio and Khavrus, Vyacheslav and Bezugly, Viktor and Ibarlucea, Bergoi and Cuniberti, Gianaurelio},
  journal={Advanced Intelligent Systems},
  volume={4},
  number={4},
  pages={2200016},
  year={2022},
  publisher={Wiley Online Library}
}

@article{huang2021_enose,
  title={Highly sensitive room temperature ammonia gas sensor using pristine graphene: The role of biocompatible stabilizer},
  author={Huang, Shirong and Panes-Ruiz, Luis Antonio and Croy, Alexander and L{\"o}ffler, Markus and Khavrus, Vyacheslav and Bezugly, Viktor and Cuniberti, Gianaurelio},
  journal={Carbon},
  volume={173},
  pages={262--270},
  year={2021},
  publisher={Elsevier}
}

@article{karnaushenko2015light_biomakers,
  title={Light weight and flexible high-performance diagnostic platform},
  author={Karnaushenko, Daniil and Ibarlucea, Bergoi and Lee, Sanghun and Lin, Gungun and Baraban, Larysa and Pregl, Sebastian and Melzer, Michael and Makarov, Denys and Weber, Walter M and Mikolajick, Thomas and others},
  journal={Adv. Healthc. Mater.},
  volume={4},
  number={10},
  pages={1517--1525},
  year={2015},
  publisher={Wiley Online Library}
}

@article{Jana2025,
author = {Jana, Dipankar and Mukherjee, Shubhrasish and Litvinov, Dmitrii and Grzeszczyk, Magdalena and Grebenchuk, Sergey and Šiškins, Makars and Gavriliuc, Virgil and Ouyang, Yihang and Chen, Changyi and Ye, Yuxuan and Yiming, Meng and Koperski, Maciej},
title = {Two-Dimensional Materials as a Multiproperty Sensing Platform},
journal = {Advanced Functional Materials},
volume = {n/a},
number = {n/a},
pages = {e16728},
  year={2025},
}

@article{Huang2023,
    author = {Huang, Shirong and Croy, Alexander and Bierling, Antonie Louise and Khavrus, Vyacheslav and Panes-Ruiz, Luis Antonio and Dianat, Arezoo and Ibarlucea, Bergoi and Cuniberti, Gianaurelio},
    title = {Machine learning-enabled graphene-based electronic olfaction sensors and their olfactory performance assessment},
    journal = {Applied Physics Reviews},
    volume = {10},
    number = {2},
    pages = {021406},
    year = {2023},
    month = {05},
    issn = {1931-9401},
}

@article{Wang2025,
author = {Wang, Wei and Chen, Li and Riemenschneider, Leif and Wang, Chen-Chen and Panes-Ruiz, Luis-Antonio and Hantusch, Martin and Chen, Yun-Xu and Zhang, Jian-Jun and Singh, Shivam and Vaynzof, Yana and L{\"o}ffler, Markus and Dianat, Arezoo and Chandrasekhar, Naisa and Huang, Shi-Rong and Cuniberti, Gianaurelio},
title = {Highly Sensitive and Selective Zinc-Based Metal–Organic Framework Derivatives Gas Sensors for Trace H2S Detection},
journal = {ACS Sensors},
volume = {10},
number = {10},
pages = {7584-7598},
year = {2025},
}

@article{Ginex2024,
title = {Quantum mechanical-based strategies in drug discovery: Finding the pace to new challenges in drug design},
journal = {Current Opinion in Structural Biology},
volume = {87},
pages = {102870},
year = {2024},
issn = {0959-440X},
author = {Tiziana Ginex and Javier Vázquez and Carolina Estarellas and F.Javier Luque},
}

@Article{Vargas2025,
author ="Vargas-Rosales, Pablo Andrés and Caflisch, Amedeo",
title  ="The physics-AI dialogue in drug design",
journal  ="RSC Med. Chem.",
year  ="2025",
volume  ="16",
issue  ="4",
pages  ="1499-1515",
publisher  ="RSC",
}

@article{Manathunga2022,
title = {Computer-aided drug design, quantum-mechanical methods for biological problems},
journal = {Current Opinion in Structural Biology},
volume = {75},
pages = {102417},
year = {2022},
issn = {0959-440X},
author = {Madushanka Manathunga and Andreas W. Götz and Kenneth M. Merz},
}

@article{Ryde2016,
author = {Ryde, Ulf and S{\"o}derhjelm, Pär},
title = {Ligand-Binding Affinity Estimates Supported by Quantum-Mechanical Methods},
journal = {Chemical Reviews},
volume = {116},
number = {9},
pages = {5520-5566},
year = {2016},
}

@article{hansen2015bob,
  title={Machine learning predictions of molecular properties: Accurate many-body potentials and nonlocality in chemical space},
  author={Hansen, Katja and Biegler, Franziska and Ramakrishnan, Raghunathan and Pronobis, Wiktor and Von Lilienfeld, O Anatole and Muller, Klaus-Robert and Tkatchenko, Alexandre},
  journal={J. Phys. Chem. Lett.},
  volume={6},
  number={12},
  pages={2326--2331},
  year={2015},
  publisher={ACS Publications}
}

@article{kovacs2025mace,
    author = {Kovács, Dávid P{\'e}ter and Moore, J. Harry and Browning, Nicholas J. and Batatia, Ilyes and Horton, Joshua T. and Pu, Yixuan and Kapil, Venkat and Witt, William C. and Magdău, Ioan-Bogdan and Cole, Daniel J. and Csányi, Gábor},
    title = {MACE-OFF: Short-Range Transferable Machine Learning Force Fields for Organic Molecules},
    journal = {Journal of the American Chemical Society},
    volume = {147},
    number = {21},
    pages = {17598-17611},
    year = {2025},
}

@article{sandip2016soap,
author ="De, Sandip and Bartók, Albert P. and Csányi, Gábor and Ceriotti, Michele",
title  ="Comparing molecules and solids across structural and alchemical space",
journal  ="Phys. Chem. Chem. Phys.",
year  ="2016",
volume  ="18",
issue  ="20",
pages  ="13754-13769",
publisher  ="The Royal Society of Chemistry",
}

@article{hourahine2020dftb,
  title={DFTB+, a software package for efficient approximate density functional theory based atomistic simulations},
  author={Hourahine, Ben and Aradi, B{\'a}lint and Blum, Volker and Bonafe, Frank and Buccheri, Alex and Camacho, Cristopher and Cevallos, Caterina and Deshaye, MY and Dumitric{\u{a}}, T and Dominguez, A and others},
  journal={J. Chem. Phys.},
  volume={152},
  number={12},
  year={2020},
  publisher={AIP Publishing}
}

@article{li2024_voc_enose,
  title={{Electronic nose for the detection and discrimination of volatile organic compounds: Application, challenges, and perspectives}},
  author={Li, Yanchen and Wang, Zike and Zhao, Tianning and Li, Hua and Jiang, Jingkun and Ye, Jianhuai},
  journal={TrAC Trends in Analytical Chemistry},
  volume={180},
  pages={117958},
  year={2024},
  publisher={Elsevier}
}

@article{wang2021artificial_sense,
  title={Artificial sense technology: emulating and extending biological senses},
  author={Wang, Jianwu and Wang, Cong and Cai, Pingqiang and Luo, Yifei and Cui, Zequn and Loh, Xian Jun and Chen, Xiaodong},
  journal={ACS nano},
  volume={15},
  number={12},
  pages={18671--18678},
  year={2021},
  publisher={ACS Publications}
}

@article{Ravera2025,
    author = {Ravera, Federico and Medrano Sandonas, Leonardo and Gutierrez, Rafael and Graziano, Mariagrazia and Cuniberti, Gianaurelio},
    title = {Are nonequilibrium effects relevant for chiral molecule discrimination?},
    journal = {J. Chem. Phys.},
    volume = {163},
    number = {1},
    pages = {014702},
    year = {2025},
    month = {07},
    issn = {0021-9606},
}

@article{MOREMLzenodo,
  doi = {10.5281/zenodo.14720508},
  url = {https://zenodo.org/records/14720508},
  year = {2025},
  author = {Chen, L. and Medrano Sandonas, L. and Huang, S. and Croy, A. and Cuniberti, G.},
  title = {MORE-QX, an extended dataset version for "MORE-Q, Dataset for molecular olfactorial receptor engineering by quantum mechanics" (Version 2.0) [Data set].},
  journal = {ZENODO}
}

\section*{Acknowledgements}
The authors gratefully acknowledge the funding by the European Union Horizon Europe EIC Pathfinder Open project “Smart Electronic Olfaction for Body Odor Diagnostics” (SMELLODI, grant agreement ID: 101046369), the Volkswagen Foundation for the Qualification Concept "Olfactorial Perceptronics" (Project ID 9B396), the DFG RTG project "B13: Neuromorphic sensing via 2D-materials-nanoparticle networks" (NeuroSense-2D, grant agreement ID: RTG2767), German Research Foundation (DFG) under the Cluster of Excellence CeTI: Centre for Tactile Internet with Human-in-the-Loop (EXC 2050), German Research Foundation (DFG) under the Cluster of Excellence REC²: Responsible Electronics in the Climate Change Era (EXC 3035), and German Research Foundation (DFG) under the Cluster of Excellence CARE: Climate-Neutral And Resource-Efficient Construction (EXC 3115), project number 533767731.
We acknowledge also the Center for Information Services and
High Performance Computing (ZIH) at TU Dresden for providing the computational resources.

\section*{Author contributions}

The work was initially conceived by LC and LMS and designed with contributions from SH, AC, and GC. LC generated the MORE-QX dataset and developed the ML regression models. The MORE-ML framework was implemented by LC and LMS, who also drafted the original manuscript. All authors discussed the results and contributed to the final version of the manuscript.

\section*{Data Availability}
MORE-QX dataset is available in a \href{https://zenodo.org/records/14720508}{ZENODO.ORG} data repository associated to this work\cite{MOREMLzenodo}.
The code and ML regression models to predict binding features within the MORE-ML framework can be found in the GitHub repository \href{https://github.com/LiC1117/MORE-Q}{MORE-Q}. 

\section*{Competing interests}
The authors declare no competing financial interests

\end{document}


\fancyhead{}
 \renewcommand{\headrulewidth}{1pt}
 \renewcommand{\footrulewidth}{1pt}
 \setlength{\arrayrulewidth}{1pt}
 \setlength{\columnsep}{6.5mm}
 \renewcommand{\figurename}{Fig.~S\!\!}
 \renewcommand{\tablename}{Table~S\!\!}

 \begin{center}
 \noindent\LARGE{Supplementary Information (SI) for:}
 \vspace{0.3cm}

 \noindent\LARGE{\textbf{Interpretable Machine Learning for Quantum-\\Informed Property Predictions in Artificial Sensing Materials}}
 \vspace{0.6cm}

\noindent\large{Li Chen\textit{$^{1}$},  
Leonardo Medrano Sandonas\textit{$^{1,}$}$^{\ast}$,
Shirong Huang\textit{$^{1}$}, 
Alexander Croy\textit{$^{2}$},
Gianaurelio Cuniberti\textit{$^{1,3,4,5,}$}$^{\ast}$},\vspace{0.5cm}

\noindent{\textit{$^{1}$~Institute for Materials Science and Max Bergmann Center of Biomaterials, TU Dresden, 01062 Dresden, Germany.}}\\[0.6em]

\noindent{\textit{$^{2}$~Institute of Physical Chemistry, Friedrich Schiller University Jena, 07737 Jena, Germany.}}\\[0.6em]

\noindent{\textit{$^{3}$~Dresden Center for Computational Materials Science (DCMS), TUD Dresden University of Technology, 01062 Dresden, Germany.}}\\[0.6em]

\noindent{\textit{$^{4}$~Cluster of Excellence CARE, TU Dresden and RWTH Aachen, Germany}}\\[0.6em]

\noindent{\textit{$^{5}$~Cluster of Excellence CeTI, TU Dresden, Germany}}\\[0.6em]

\noindent{$^{\ast}$~Corresponding author: Leonardo Medrano Sandonas (\texttt{leonardo.medrano@tu-dresden.de}), Gianaurelio Cuniberti (\texttt{gianaurelio.cuniberti@tu-dresden.de})}
\end{center}
 
\vspace{0.5in}
\clearpage
\section{Property and abbreviation tables}
\begin{table}[ht]
\centering
\caption{List of abbreviations used in the manuscript.}

\begin{tabular}{ll}
\hline
\textbf{Abbreviation} & \textbf{Definition} \\ 
\hline
BOV    & Body odor volatilomes \\
OM     & BOV molecule \\
REC    & Receptor \\
DM     & BOV–receptor dimer \\
BD     & Binding features \\
OM–REC & BOV–receptor complex \\
CPLX   & BOV–receptor–surface complex \\
SUB    & Receptor–surface substrate system \\
TR     & Training set \\
TE     & Test set \\
UMAP   & Uniform Manifold Approximation and Projection for Dimension Reduction \\
t-SNE  & t-distributed Stochastic Neighbor Embedding \\
RF     & Random Forest \\
GB     & Gradient Boosting Decision Tree \\
CAT    & CatBoost \\
XGB    & XGBoost \\
LGBM   & LightGBM \\
HOMO   & Highest Occupied Molecular Orbital \\
LUMO   & Lowest Unoccupied Molecular Orbital \\
SHAP   & SHapley Additive exPlanations \\
MORE-Q & Molecular Olfactorial Receptor Engineering by Quantum Mechanics \\
MORE-QX & Extended Molecular Olfactorial Receptor Engineering by Quantum Mechanics \\
\hline
\end{tabular}
\label{abbreviations}
\end{table}
\clearpage

\begin{table}[h]
  \centering
  \caption{List of the Quantum-mechanical (QM) properties (and corresponding symbols) taken from MORE-QX dataset analyzed in this work. In the units provided for each of these QM properties, $a_0$ stands for the atomic unit of length (Bohr radius). Property types are classed according to the building blocks as follow: Monomer (OM, REC), Dimer(DM), complex system (CPLX), and binding feature (BD). A full characterization of QM properties in MORE-QX dataset can be found in the MORE-Q manuscript\cite{more-q}.}
  \begin{tabular}{c c c c}
    \toprule
    Symbol & Property description & Units  & Type \\ \midrule
    $\mu_{\rm z, OM}$  & OM dipole moment $z$ component & Debye & Monomer \\ 
    $\epsilon_{\rm H, REC}$  & REC HOMO orbital energy & eV  & Monomer \\ 
    $\epsilon_{\rm H, OM}$  & OM HOMO orbital energy & eV  & Monomer \\
    $\mu_{\rm REC}$  & REC total dipole moment & Debye & Monomer \\ 
    $\mu_{\rm OM}$  & OM scaler total & Debye  & Monomer \\ 
    $Q_{xy, \rm REC}$  & REC quadrupole moment tensor $xy$ component & Buckingham & Monomer  \\
    $I_{xy,\rm REC}$  & Inertia moment tensor $xy$ component & $\mathrm{amu}\cdot\mathrm{\AA}$ & Monomer\\ 
    $\epsilon_{\rm H, DM}$  & OM-REC HOMO orbital energy & eV  & Dimer \\ 
    $\epsilon_{\rm L, DM}$  & OM-REC LUMO orbital energy & eV  & Dimer \\ 
    $\mu_{z,\rm DM}$  & OM-REC dipole moment $z$ component  & Debye  & Dimer \\
    $\alpha_{\rm s,DM}$  & OM-REC molecular isotropic polarizability & $a_0^3$ & Dimer \\ 
    $\mu_{\rm DM}$  & OM-REC total dipole moment  & Debye & Dimer \\ 
    $\epsilon_{\rm gap,DM}$  & OM-REC HOMO-LUMO gap & eV & Dimer \\
     $E_{\rm int}$  & OM-REC binding energy & eV & Dimer \\ 
     $E_{f, \rm CPLX}$  & OM-REC-graphene Fermi level & eV & Complex \\ 
    $E_{\rm ads}$  & Adsorption energy & eV  & Binding feature \\ 
    $\Delta\phi$  & Work function change & eV & Binding feature \\
     $\Delta Q$  & Bader charge transfer & e & Binding feature \\ 
    \bottomrule
  \end{tabular}
  \label{tab:sample_table}
\end{table}

\clearpage
\begin{table}[h]
  \centering
  \caption{Full list of the Quantum-mechanical (QM) properties (and corresponding symbols) used as input electronic features $D_{\rm ele}$. The units and property type categories provided are the same as those in Tab. S2. Property types are classed according to the building blocks as follow: Monomer (OM, REC), dimer (DM), complex system (CPLX), and binding feature (BD). One property might simultaneously apply to different systems. As a result, 130 features are used as the original features for machine learning models. A full characterization of QM properties in MORE-QX dataset can be found in the MORE-Q manuscript\cite{more-q}.}
    \begin{tabular}{llllcll}
    \toprule
       \#& Property & Symbol & Unit & Dimension & System & HDF5 keys\\
       \hline
        1 & Total PBE+D3 energy & $E_{\rm tot}$ &eV & 1 & OM, REC, DM & 'ePBE+D3'  \\
        2 & Nuclear repulsion energy &$E_{\rm nuc}$ & eV & 1 &OM, REC, DM& 'eNUC' \\
        3 & Electronic repulsion energy & $E_{\rm ele}$& eV & 1 &OM, REC, DM & 'eELE'  \\
        4 & One electron energy & $E_{\rm 1e}$&eV & 1 & OM, REC, DM& 'e1E'  \\
        5 & Two electron energy & $E_{\rm 2e}$&eV & 1 & OM, REC, DM& 'e2E'  \\
        6 & Virial potential energy & $E_{\rm pe}$ & eV & 1 & OM, REC, DM & 'ePE'  \\
        7 & Virial kinetic energy & $E_{\rm ke}$ &eV & 1 & OM, REC, DM& 'eKE'  \\
        8 & Exchange energy & $E_{\rm x}$ &eV & 1 &OM, REC, DM& 'eX'  \\
        9 & Correlation energy & $E_{\rm c}$ &eV & 1 & OM, REC, DM& 'eC'  \\
        10 & Exchange-correlation energy & $E_{\rm xc}$ & eV & 1 & OM, REC, DM& 'eXC'  \\
        11 & Total D3 energy & $E_{\rm D3}$ &eV & 1 & OM, REC, DM& 'eD3'  \\
        12 & Dispersion E6 energy & $E_{\rm 6}$ &eV & 1 & OM, REC, DM& 'eE6'  \\
        13 & Dispersion E8 energy & $E_{\rm 8}$ &eV & 1 & OM, REC, DM& 'eE8' \\
        14 & HOMO energy & $\epsilon_{\rm H}$ & eV & 1 & OM, REC, DM& 'eH'  \\
        15 & LUMO energy & $\epsilon_{\rm L}$ & eV & 1 & OM, REC, DM& 'eL'  \\
        16 & HOMO-LUMO gap & $\epsilon_{\rm gap}$ & eV & 1 & OM, REC, DM& 'HLgap'  \\
        17 & Isotropic molecular $C_6$ coefficient &$C_6$ & $E_{\mathrm{h}}\cdot a_0^6$ & 1 & OM, REC, DM & 'mC6'  \\
        18 & Total dipole moment & $\mu$ & D & 3 & OM, REC, DM & 'vDIP' \\
        19 & Scalar total dipole moment & $\mu_{\rm s}$ &D & 1 & OM, REC, DM & 'DIP'  \\
        20 & Rotational spectrum constant & $B$ & MHz & 3 &  OM, REC, DM & 'vRS'  \\
        21 & Rotational dipole moment  & $\mu_{\rm B}$ & d & 3 & OM, REC, DM & 'vRSDIP'  \\
        22 & Total quadrupole moment tensor& $Q$ & Buckingham & 6 & OM, REC, DM & 'TQP'  \\
        23 & Isotropic molecular quadrupole  & $Q_{\rm s}$ & Buckingham & 1 & OM, REC, DM & 'mQP'  \\
        24 & Molecular polarizabillity tensor & $\alpha$ & $a_0^3$  & 6 & OM, REC, DM & 'mTPOL'  \\
        25 & Molecular isotropic polarizability & $\alpha_{\rm s} $ & $a_0^3$ & 1 & OM, REC, DM & 'mPOL'  \\
        26 & Radius of gyration & $R_{\rm g}$  & \AA & 1 & OM, REC, DM & 'RG'  \\
        27 & Inertia moment tensor & $I_{\rm TS}$ & amu$\cdot$\AA$^2$ & 6 & OM, REC, DM  & 'IM' \\
        28 & Atomisation energy &$E_{\rm at}$ & eV & 1 & OM, REC, DM & 'eAT'  \\
        29 & Binding energy & $E_{\rm int}$ & eV & 1 & DM & 'eBIND'\\
    \bottomrule
  \end{tabular}
  
  \label{tab:sample_table}
\end{table}

\clearpage
\section{MORE-QX data distribution}

\begin{figure}[h]
    \centering
    \includegraphics[width=1\linewidth]{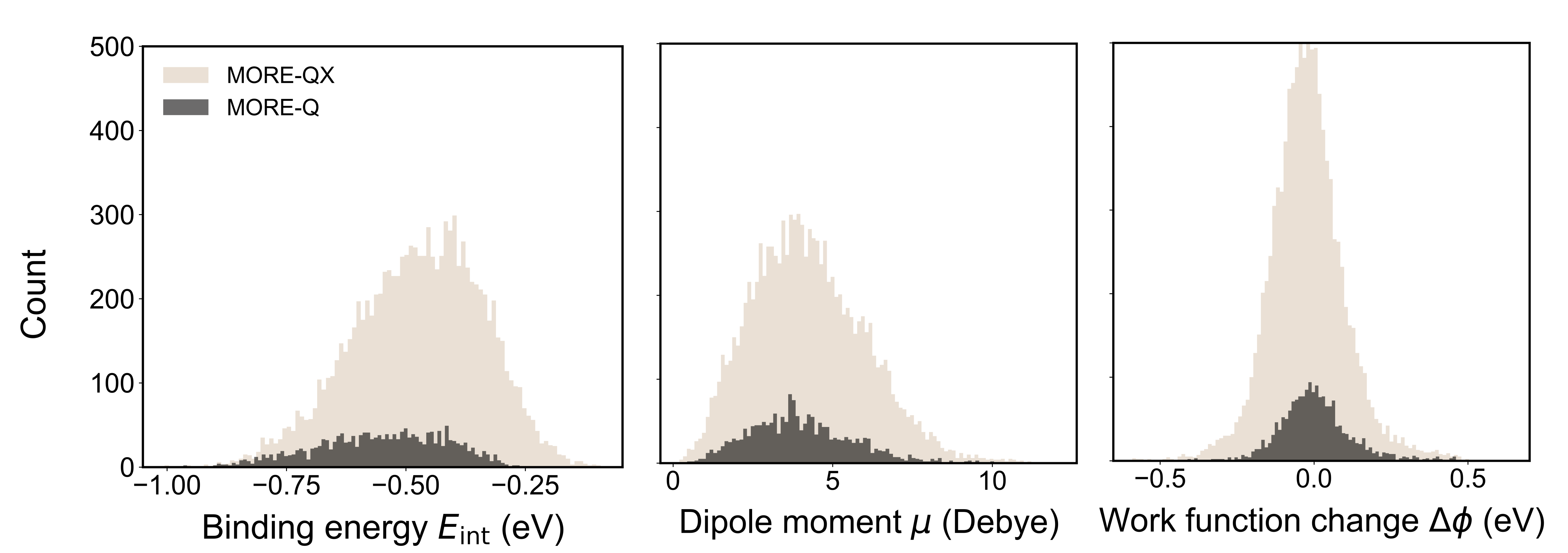}
    \caption{Three examples for the property distribution comparison between MORE-QX (brown) and MORE-Q (black)}
    \label{prop_dist}
\end{figure}

\begin{figure}[h]
    \centering
    \includegraphics[width=0.5\linewidth]{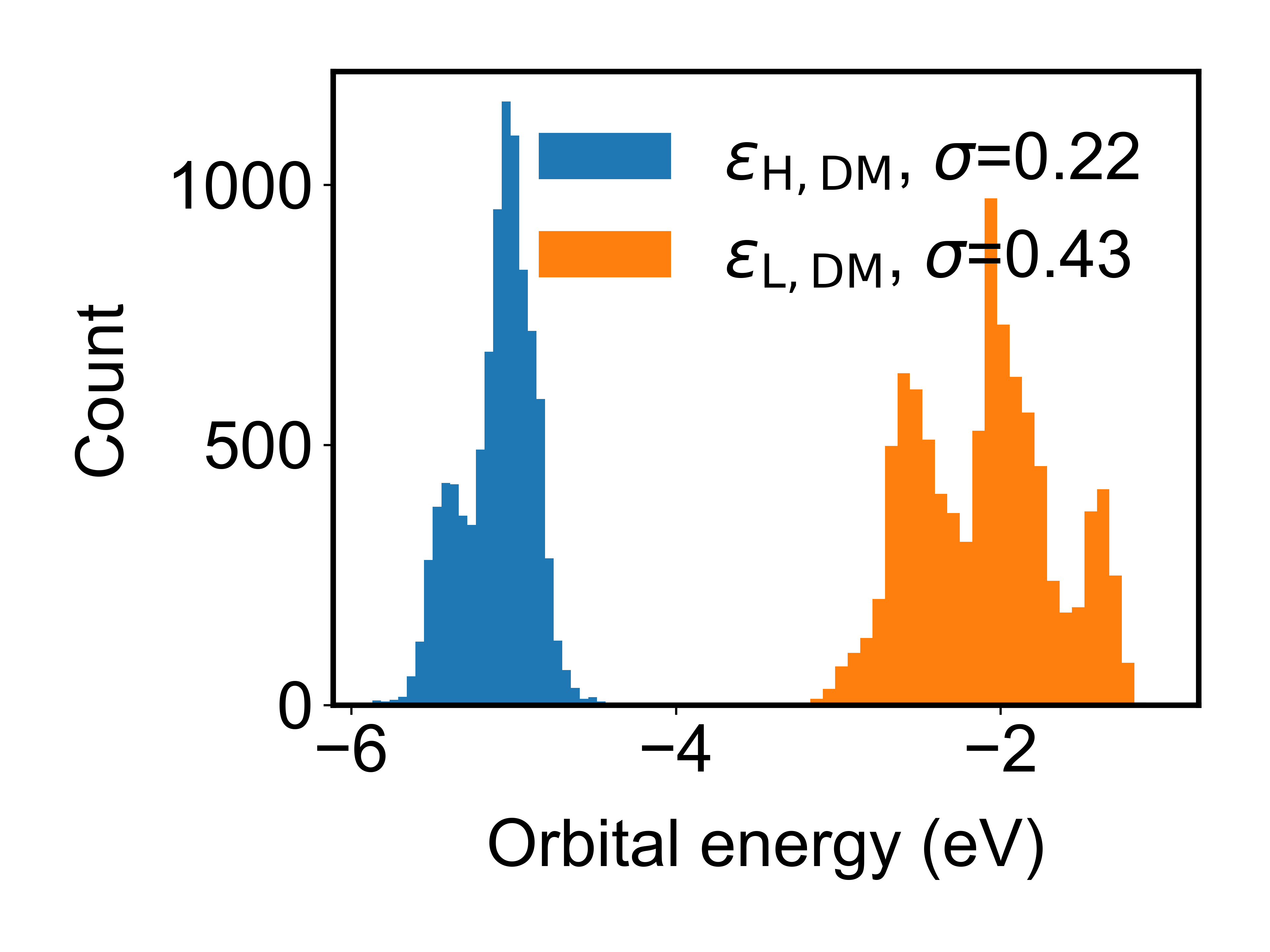}
    \caption{The distribution of the dimer HOMO ($\epsilon_{\rm H,DM}$, blue) and LUMO  ($\epsilon_{\rm L,DM}$, orange) orbital energies. We show the respective variances $\sigma$.}
    \label{HL_distribution}
\end{figure}

\clearpage
\section{Tree-based Machine learning models}
\begin{figure}[h]
    \centering
    \includegraphics[width=0.9\linewidth]{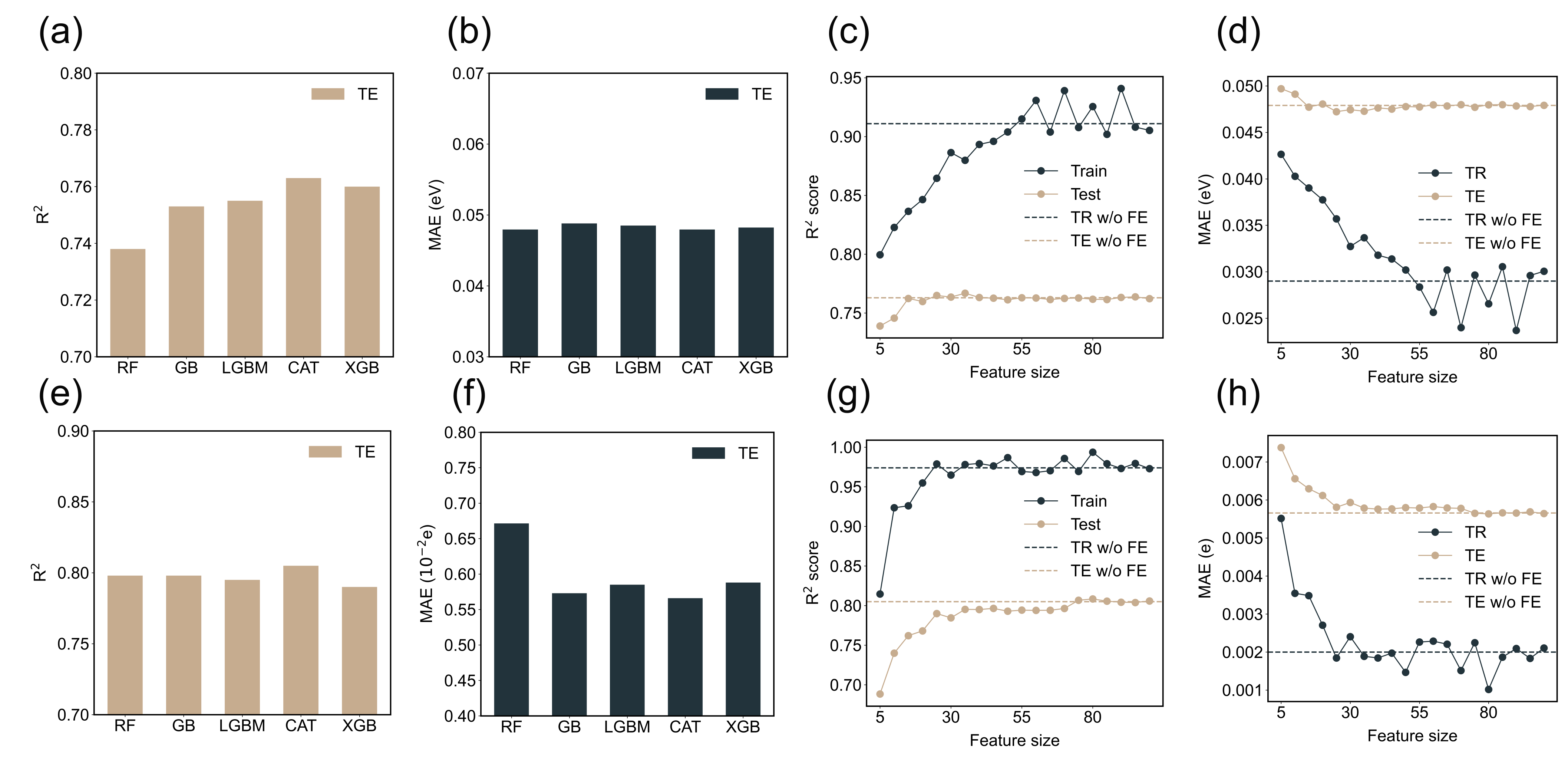}
    \caption{Model benchmark for work function change $\Delta \phi$ (a)$\sim$(d) and charge transfer $\Delta Q$ (e)$\sim$(h). The best model for both binding features remains Catboost (CAT). And the feature size for $\Delta \phi$ and $\Delta Q$ accounts for 60 and 80.}
    \label{Benchmark_FE}
\end{figure}

Tree-based ML models, which belongs to the ensemble learning category, are divided into bagging and gradient boosting methods, and their performance towards different regression tasks were benchmarked in Fig. S\ref{Benchmark_FE}. In this section, we introduce the main features for each model regarding regression task.
The uniform definition is given as follow:
\begin{equation}
    F(x) \;=\; \sum_{m=1}^M w_m\,h_m(x;\,\theta_m),
    \label{tree-tot}
\end{equation}
where $h_m(x;\theta_m)$  is the $m^{\mathrm{th}}$ tree (weak learner) and its structure and leaf values are determined by the hyperparameter $\theta_m$. And the $w_m$ is the weight of the $m^{\mathrm{th}}$. 
The objective function for training process and the loss function for each tree is defined as: 
\begin{equation}
\theta_m = \mathop{\arg\min}_{\theta}\;\sum_{i=1}^n \ell\bigl(y_i,\;F_{m-1}(x_i) + w_m\,h_m(x_i;\theta)\bigr)
\;+\;\Omega\bigl(h_m(x;\theta)\bigr), 
\label{tree-loss}
\end{equation}
and output of the model is then updated by:
\begin{equation}
F_m(x) = F_{m-1}(x) + w_m\,h_m(x;\theta_m),
\quad
F_0(x)=\bar y,
\label{next-tree}
\end{equation}
where $\ell(y,\hat y)$ is the loss function bewteen ground truth and prediction value and $\Omega(h)$ denotes the complexity of the tree $h$ and regularizes the training process. And the initial residual \ie output of the $0^{\mathrm{th}}$ results are set to be average of the output $\bar y$.

\subsubsection*{Random forest (RF)}
Random forest trains weak learners independently by simply averaging the predictions from all $M$ the individual trees which turns $w_m$ into $\frac{1}{M}$ in Eq. S\ref{tree-tot}.
And the model complexity is controlled via hyperparamters such as tree depth, minimum samples per leaf. An explicit regularization term $\Omega(h)$ is typically omitted. 

\subsubsection*{Gradient boosting decision tree (GB)}
In GB method, a constant learning rate is allocated to $w_m = \nu \in (0,1]$. In each training iteration, pseudo-residuals are computed as following:
\begin{equation}
    r_{im} = - \frac{\partial\ell(y_i,F_{m-1}(x_i))}{\partial F_{m-1}(x_i)}
\end{equation}
And new tree $h_m{x}$ is fitted to the residuals, and then the model is updated via Eq. S\ref{next-tree}. The regularization term is explicitly given as: 
\begin{equation}
    \Omega(h) = \gamma T + \frac{1}{2}\lambda\Sigma_j w^2_j,
\end{equation}
where $T$ denotes the number of leaves, and $\gamma$, $\lambda$ penalize complexity of the tree structre and leaf weights.

\subsubsection*{XGBoost (XGB)}
To enhance training speed and also model stability, the loss function part Eq. S\ref{tree-loss} is modified by Taylor expansion at the $m^{\mathrm{th}}$ prediction for training sample $i$ and hence Eq. S\ref{tree-loss} turns into:
\begin{equation}
    \theta_m = \mathop{\arg\min}_{\theta}\;\sum_{i=1}^n \bigl(g_i h(x_i;\theta)+ \frac{1}{2}h_ih(x_i)^2 \bigr)
\;+\;\Omega\bigl(h_m(x;\theta)\bigr), 
\end{equation}
where $g_i = \frac{\partial \ell(y_i,\hat{y})}{\partial \hat{y}}\big|_{\hat{y}=F_{m-1}(x_i)}$ and $h_i = \frac{\partial^2 \ell(y_i,\hat{y})}{\partial^2 \hat{y}}\big|_{\hat{y}=F_{m-1}(x_i)}$ are the first and second derivative (gradient and hessian) of the loss $\ell$ at the $m^{\mathrm{th}}$ prediction.

\subsubsection*{LightGBM (LGBM)\& Catboost (CAT)}
Built on XGBoost, LGB adapts the histogram and leaf-wise tree growing strategy to improve training speed with fewer memory counts, while CAT employs the oblivious tree by forcing the ical feature on the node splitting within each layer to overcome overfitting problems further. 

\clearpage

\section{Additional results on "Freedom of design" in the binding feature space}

\begin{figure}[!ht]
    \centering
    \includegraphics[width=0.6\linewidth]{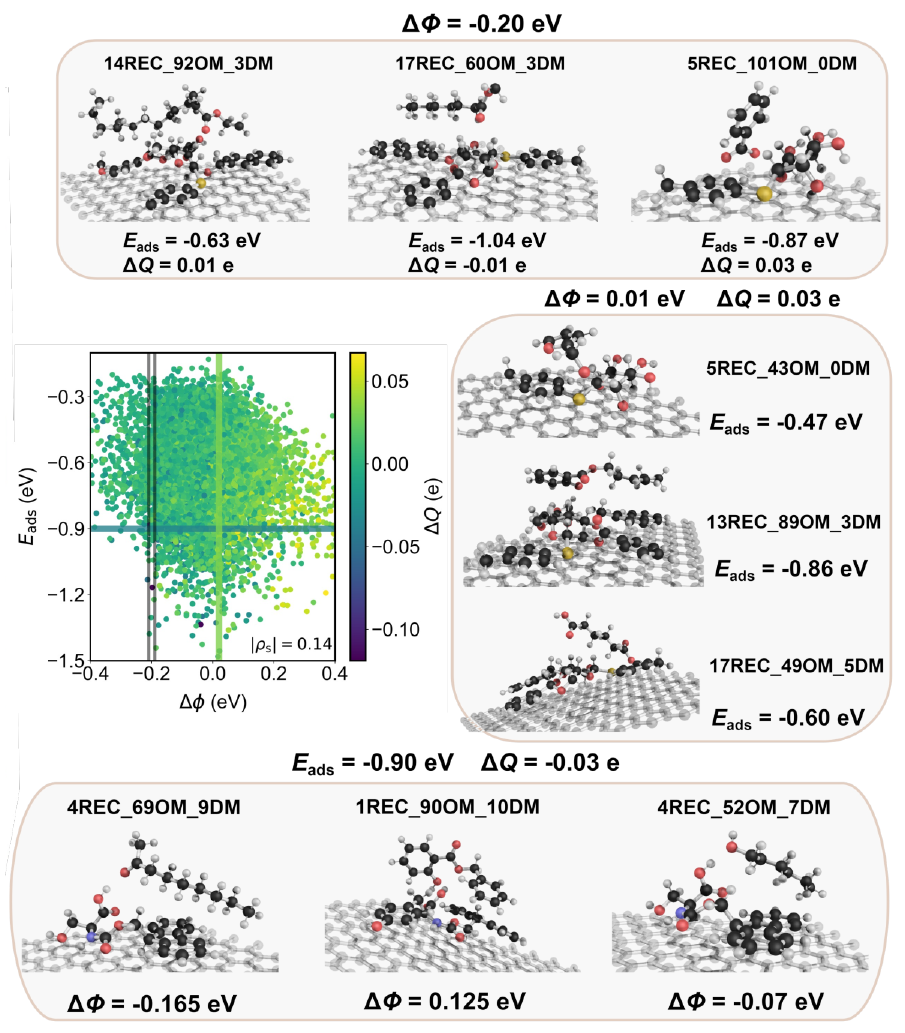}
    \caption{Examples for design tasks under the freedom of design conjecture. Each system is named by \textit{n}REC\_\textit{m}OM\_\textit{l}DM, where \textit{n}, \textit{m} and \textit{l} refer to the number of receptor, BOV molecules and their dimer conformer. The full list of BOV and receptor molecules and their number can be viewed in MORE-Q \cite{more-q}. Top panel: design task for constraining $\Delta \phi = -0.2\,\rm \pm 0.01\,eV$. Right panel: design task for constraining $\Delta \phi = 0.01\,\rm \pm 0.01\,eV$. Bottom panel: constraining the $E_{\rm ads} = -0.9\,\rm \pm 0.01\,eV$ and $\Delta Q = -0.03\,\rm \pm 0.001\,e$. Middle panel: scatter plot between adsorption energy and work function change, colored by charge transfer.}
    \label{FOD}
\end{figure}

The scatter plot in the middle panel of Fig. S\ref{FOD} illustrates the correlation between adsorption energy and work function change, yielding a correlation coefficient of $|\rho_{\rm s}| = 0.14$. This very weak correlation provides an ideal example for exploring the freedom of design conjecture.
Therefore, we start firstly with a simple constraint design task given only $\Delta \phi = -0.20\, \rm \pm 0.01\,eV$, in which corresponding the $50^{\rm th}$ of the negative half distribution of the $\Delta \phi$, 
as shown in the unfilled dark lines in the scattering plot of Fig. S\ref{FOD}. 
Along the dark lines, the $E_{\rm ads}$ varies in a good range roughly from $-0.3$ and $-1.1\, \rm eV$, whose value might be correlating to the DM interacting area especially in weak interaction systems driven by electrostatics or Van der Waals interaction \cite{kreuzer2012physisorption}. 
Contrary to the three systems highlighted in the top panel of Fig. S\ref{FOD}, each satisfying identical $\Delta \phi$, display markedly different adsorption energies: the largest OM (ID 92) exhibits $E_{\rm ads} = -0.63\,\rm eV$, while the other two denotes $-1.04$ and $-0.87\, \rm eV$ with smaller molecular size. 
The deviation in $E_{\rm ads}$ scaling law might be ascribed to the $\rm O$-containing pocket formation in DM interaction on the surface.   
Besides, these DM interaction yields the almost the same $\Delta \phi$ and different $\Delta Q$ in both value and sign manifesting the freedom of design conjecture in finding complex structures with low-correlated $E_{\rm ads}$ and $\Delta Q$ under one simple $\Delta \phi$ constrain and also reflecting the complexity in correlating the $\Delta \phi$ to the morphological and compositional aspects of the systems.
Next, we impose more stringent design constraints by targeting $\Delta \phi = 0.01 \pm 0.01\,\rm eV$ \ie $\Delta \phi$ non-dominant cases and relative larger $\Delta Q = 0.03\, \rm eV$ as shown in the yellow strip in the scattering plot.
Under more stringent conditions, as shown in the middle panel, we can still identify systems with tailored $E_{\rm ads}$. In these cases, the scaling law holds from 5REC–43OM over 13REC–89OM to 17REC–49OM.
As a final demonstration, we impose constraints of $E_{\rm ads} = -0.9\,\rm \pm 0.01 eV$ and $\Delta Q = -0.03 \pm 0.001 \rm e$, thereby ensuring identical recovery times and a charge‐transfer–dominant mechanism, as indicated by the grey horizontal line in the bottom panel. 
In these complexes, hydrogen bonding between the OM and REC molecules compensates for variations in interaction‐area size, allowing long-chain, pyrene‐ring-based, and small-size systems to exhibit identical $E_{\rm ads}$ values.
Interestingly, the three systems yield $\Delta \phi$ with both different values and sign under an identical $\Delta Q$, which manifests again the freedom of design again from another perspective. 

\clearpage
\section{The entire workflow for MORE-ML}
\begin{figure}[h]
    \centering
    \includegraphics[width=1\linewidth]{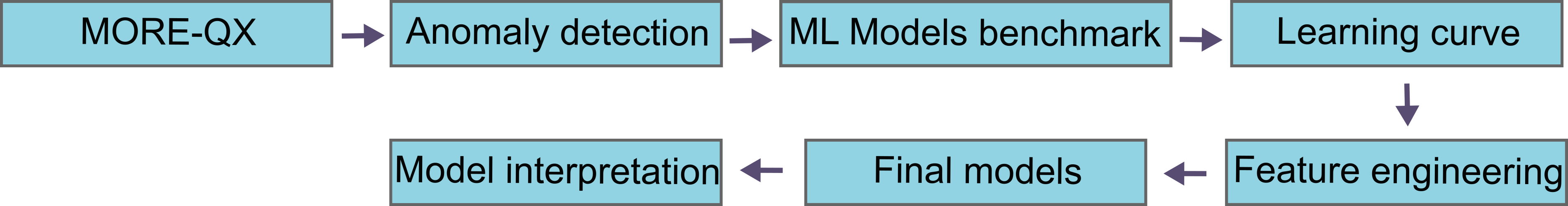}
    \caption{Overview of the entire Machine learning workflow in MORE-ML. } 
    \label{SI_ML_WORKFLOW}
\end{figure}

\clearpage
\section{Anomaly detection}
\begin{figure}[h]
    \centering
    \includegraphics[width=1\linewidth]{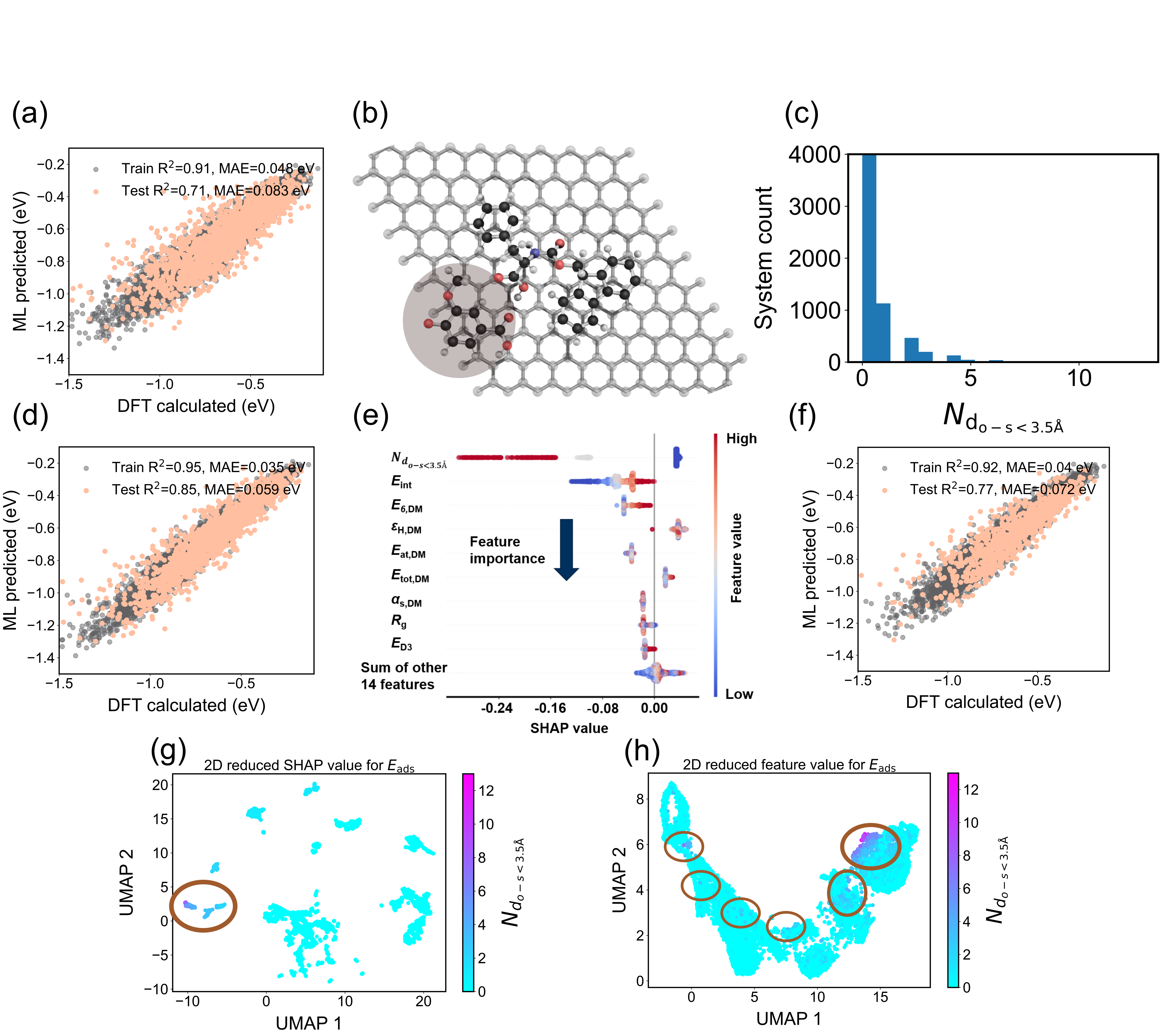}
    \caption{Anomaly detection workflow for MORE-QX. (a) Adsorption energy $E_{\rm ads}$ prediction using the dimer properties. (b) Atomistic illustration for an anomaly case, where the OM is exposed mainly to graphene. (c) The distribution of the $N_{d_{o-s<3.5 \text{\AA}}}$ among the $10,411$ systems. (d) $E_{\rm ads}$ prediction by adding $N_{d_{o-s<3.5 \text{\AA}}}$ into the input features under the identical hyperparameters. 
    (e) SHAP analysis beewarms plot from the prediction results in (d). (f) The $E_{\rm ads}$ prediction results after removing the $932$ outliers. (g) UMAP plot for clustering the outliers using SHAP value. (h) UMAP plot for clustering the outliers using feature value. The green circles in (g) and (h) are highlighting the location of the outliers.}
    \label{SI_AD}
\end{figure}

During our initial benchmarking runs for predicting adsorption energy ($E_{\rm ads}$) with XGBoost (XGB), performance remained unsatisfactory after multiple trials, as shown in Fig. S\ref{SI_AD} (a). Then we checked the geometries of the outliers on Fig. S\ref{SI_AD} (a). 
In the example shown in Fig. S\ref{SI_AD} (b), the OM’s dominant interaction is with graphene rather than the receptor. 
Accordingly, for each system we counted the number of atoms whose distance to the surface is $>3.5\text{\AA}$, which is $\pi-\pi$ stacking distance and defined this quantity as the descriptor $N_{d_{o-s < 3.5\text{\AA}}}$. 
The distribution of $N_{d_{o-s < 3.5 \text{\AA}}}$ are depicted in Fig. S\ref{SI_AD} (c) indicating that there are indeed minor exceptional systems which have unignorable atoms on the OM interacting mainly graphene.
The distribution of $N_{d_{o-s <3.5\text{\AA}}}$ shown in Fig. S\ref{SI_AD} (c) reveals a small subset of exceptional systems in which a non-negligible number of OM atoms interact primarily with graphene. 
To identify the anomalies most responsible for degrading model performance, we add $N_{d_{o-s < 3.5 \text{\AA}}}$ as a 'diagnostic descriptor' into the input feature.
As shown in Fig. S\ref{SI_AD} (d), performance improves substantially relative to Fig. S\ref{SI_AD} (a), indicating that inclusion of $N_{d_{o-s < 3.5\text{\AA}}}$ could help tree-based model better classify the data points associated with the new diagnostic descriptor.
Therefore, $N_{d_{o-s <3.5\text{\AA}}}$ is highly informative and would gain much importance in predicting $E_{\rm ads}$. SHAP analysis (Fig. S\ref{SI_AD} (e)) corroborates this, with $N_{d_{o-s < 3.5 \text{\AA}}}$ ranking as the most important feature. 
Interestingly, although most systems with low $N_{d_{o-s < 3.5 \text{\AA}}}$ contribute only marginally to the model, a subset exerts a disproportionately large influence on the predictions \eg red points. 
Next, we examined the clustering of these systems to identify the outliers, on which the $N_{d_{o-s <3.5 \text{\AA}}}$ is the sole anomalous factor.
To this end, we used UMAP for dimensionality reduction because it preserves global structure relevant to cluster formation. Moreover, we embedded SHAP values rather than raw feature values, since SHAP value captures each feature’s contribution and provides a more discriminative representation, allowing samples with similar contribution profiles to cluster more clearly.
As highlighted in Fig. S\ref{SI_AD} (g), the major outliers with high $N_{d_{o-s <3.5 \text{\AA}}}$ cluster in neighboring regions, whereas points with high $N_{d_{o-s3.5 < \text{\AA}}}$ in Fig. S\ref{SI_AD} (h) are distributed broadly and do not exhibit a shared similarity structure in the UMAP space based feature value. Therefore, the outliers' SHAP values form a better cluster than the feature values.
Therefore, we identified the $932$ highlighted data points in Fig. S\ref{SI_AD} (g) as the anomalies and removed them from our dataset, as they do not contribute to the OM-REC interaction and hence are not significant for the receptor design tasks.
In addition, we list the hyperparameter table of XGBoost for reproduction purposes. 

\begin{table}[h]
  \centering
  \caption{XGBoost hyperparameter list used for Anomaly detection.}

  \begin{tabular}{l l}
    \toprule
    Hyperparameter & Value \\
    \midrule
    \texttt{lambda}                & 0.603225906844846 \\
    \texttt{alpha}                 & 0.007555374299051771 \\
    \texttt{colsample\_bytree} & 0.6000000000000001 \\
    \texttt{subsample}       & 0.9 \\
    \texttt{learning\_rate}  & 0.016 \\
    \texttt{n\_estimators}   & 2000 \\
    \texttt{max\_depth}      & 13 \\
    \texttt{min\_child\_weight} & 62 \\
    \texttt{random\_state}   & 20240815 \\
    \texttt{n\_jobs}         & 1 \\
    \bottomrule
  \end{tabular}
  \label{xgb-best-params}
\end{table}

\clearpage
\section{Learning curve}
\begin{figure}[h]
    \centering
    \includegraphics[width=0.7\linewidth]{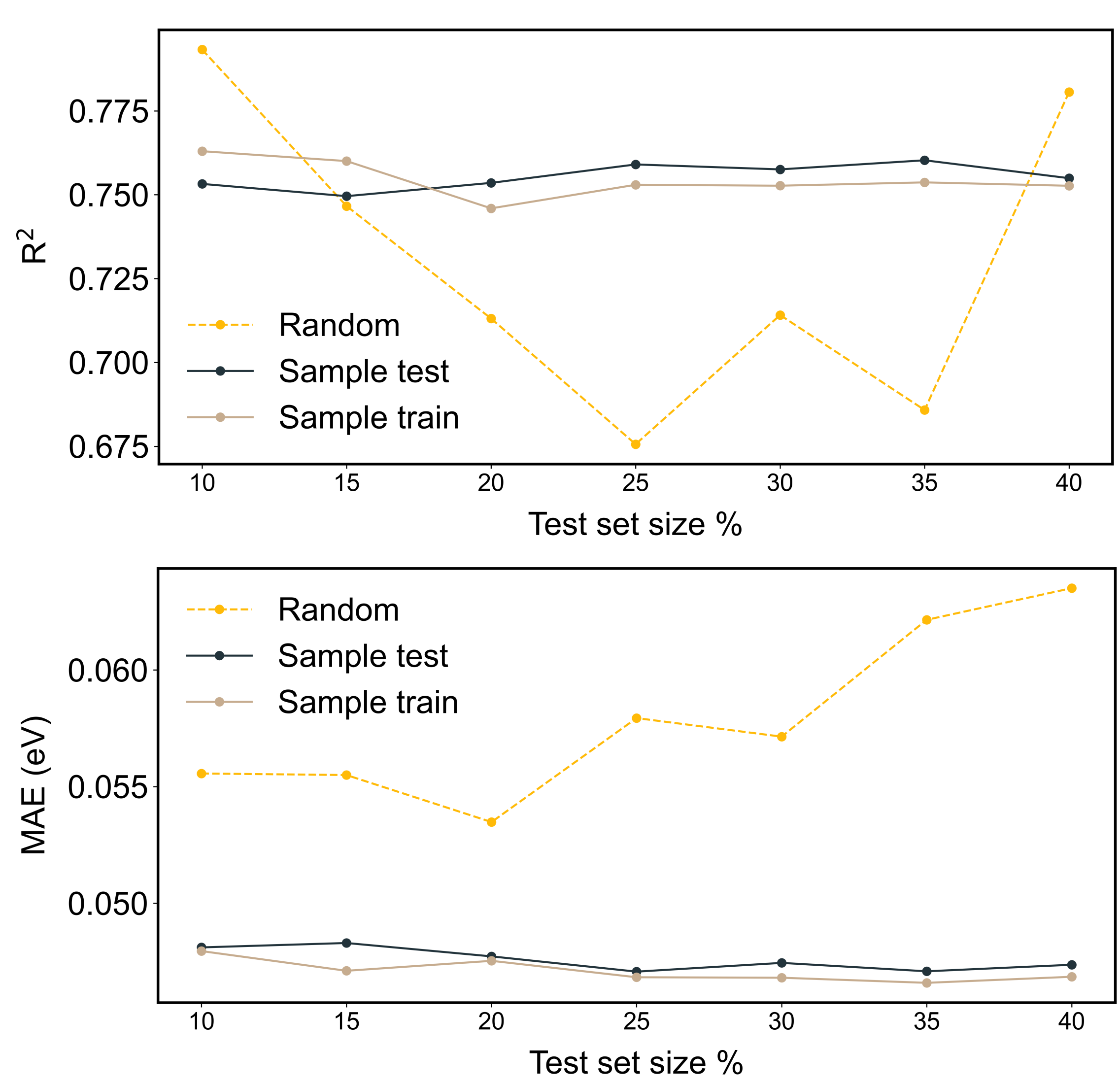}
    \caption{Learning curve for work function change $\Delta \phi$ prediction with the test set ratio varying from $10\%\sim40\%$ in the whole dataset by Catboost (CAT). Top panel: the R-square score. Bottom panel: the mean absolute error (MAE). The yellow dashed line was generated using random splitting. The black and green dot-line were generated using farthest-point-sampling method by sampling train set (black) and test set (green) on the t-SNE space. For every point, the model's hyperparamters were optimized by Bayesian optimization.}
    \label{SI_Learning_Curve}
\end{figure}

To obtain the best train-test ratio, we conducted a learning curve study, as shown in Fig. S\ref{SI_Learning_Curve}. Firstly, we can observe the model's performance stability using FPS methods for sampling compared to random splitting (varying from the test data size and no guarantee to the distribution similarity between the train and test set), as the FPS ensures a homogeneous-distribution sampling between the sampled and source dataset. Therefore, we chose the FPS as our splitting method.
Secondly, by using FPS, we can choose either the train or test set, and the counterpart is the remaining after sampling.
We noticed that this tiny difference would lead to a slight performance discrepancy.
As shown in Fig. S\ref{SI_Learning_Curve}, the performance selecting the test set is generally slightly worse than selecting the train set, as selecting the most representative test set also indicates selecting the most challenging test set.
Therefore, concerning our imbalanced dataset and the model performance, we select the splitting ratio $9:1$ and the train set. 

\clearpage
\section{Hyperparameter search space}
\begin{table}[htbp]
\centering
\caption{Hyperparameter search space of the tree models for Bayesian optimization used in this work.}
\scriptsize  %
\begin{tabular}{llp{5cm}}
\toprule
Model & Hyperparameter & Search Space \\
\midrule
\multirow{7}{*}{\rotatebox{90}{XGBoost}} 
& lambda, alpha & $[10^{-3}, 10^{-2}, ..., 10.0]$ (Log-uniform) \\
& colsample\_bytree, subsample & $[0.1, 0.2, \ldots, 1.0]$ \\
& learning\_rate & $[0.008, 0.010, \ldots, 0.020]$ \\
& n\_estimators & $[500, 1000, 3000, 5000, 7000]$ \\
& max\_depth & $[2, 4, 6, 8, 10, 12, 14]$ \\
& min\_child\_weight & $[1,2,...,300]$ (Integer) \\
\midrule
\multirow{5}{*}{\rotatebox{90}{RF}} 
& n\_estimators & $[500, 1000, 3000, 5000, 7000]$ \\
& max\_depth & $[2, 4, 6, 8, 10, 12, 14]$ \\
& min\_samples\_split/leaf & $[2, 4, 6, 8, 10, 12, 14, 16, 18, 20]$ \\
& max\_features & $[0.1, 1.0]$ (Float) \\
\midrule
\multirow{6}{*}{\rotatebox{90}{GB}} 
& n\_estimators & $[500, 1000, 3000, 5000, 7000]$ \\
& learning\_rate & $[0.008, 0.010, \ldots, 0.020]$ \\
& max\_depth & $[2, 4, 6, 8, 10, 12, 14]$ \\
& min\_samples\_split/leaf & $[2, 4, 6, 8, 10, 12, 14, 16, 18, 20]$ \\
& subsample & $[0.1, 0.2, \ldots, 1.0]$ \\
\midrule
\multirow{7}{*}{\rotatebox{90}{LightGBM}} 
& num\_leaves & $[20, 150]$ (Integer) \\
& max\_depth & $[2, 4, 6, 8, 10, 12, 14]$ \\
& learning\_rate & $[0.008, 0.010, \ldots, 0.020]$ \\
& n\_estimators & $[500, 1000, 3000, 5000, 7000]$ \\
& min\_child\_samples & $[5, 50]$ (Integer) \\
& subsample, colsample\_bytree & $[0.1, 0.2, \ldots, 1.0]$ \\
\midrule
\multirow{5}{*}{\rotatebox{90}{CatBoost}} 
& iterations & $[500, 1000, 3000, 5000, 7000]$ \\
& learning\_rate & $[0.008, 0.010, \ldots, 0.020]$ \\
& depth & $[2, 14]$ (Integer) \\
& l2\_leaf\_reg & $[1.0, 10.0]$ (Float) \\
& border\_count & $[32, 255]$ (Integer) \\
\bottomrule
\centering
\end{tabular}

\label{tab:hyperparams}
\end{table}

\clearpage
\section{Model benchmark information}

\clearpage
\section{Performance of the geometrical descriptor $D_{\rm geo}$ and Mulliken atomic charge $q$}
\begin{figure}
    \centering
    \includegraphics[width=1\linewidth]{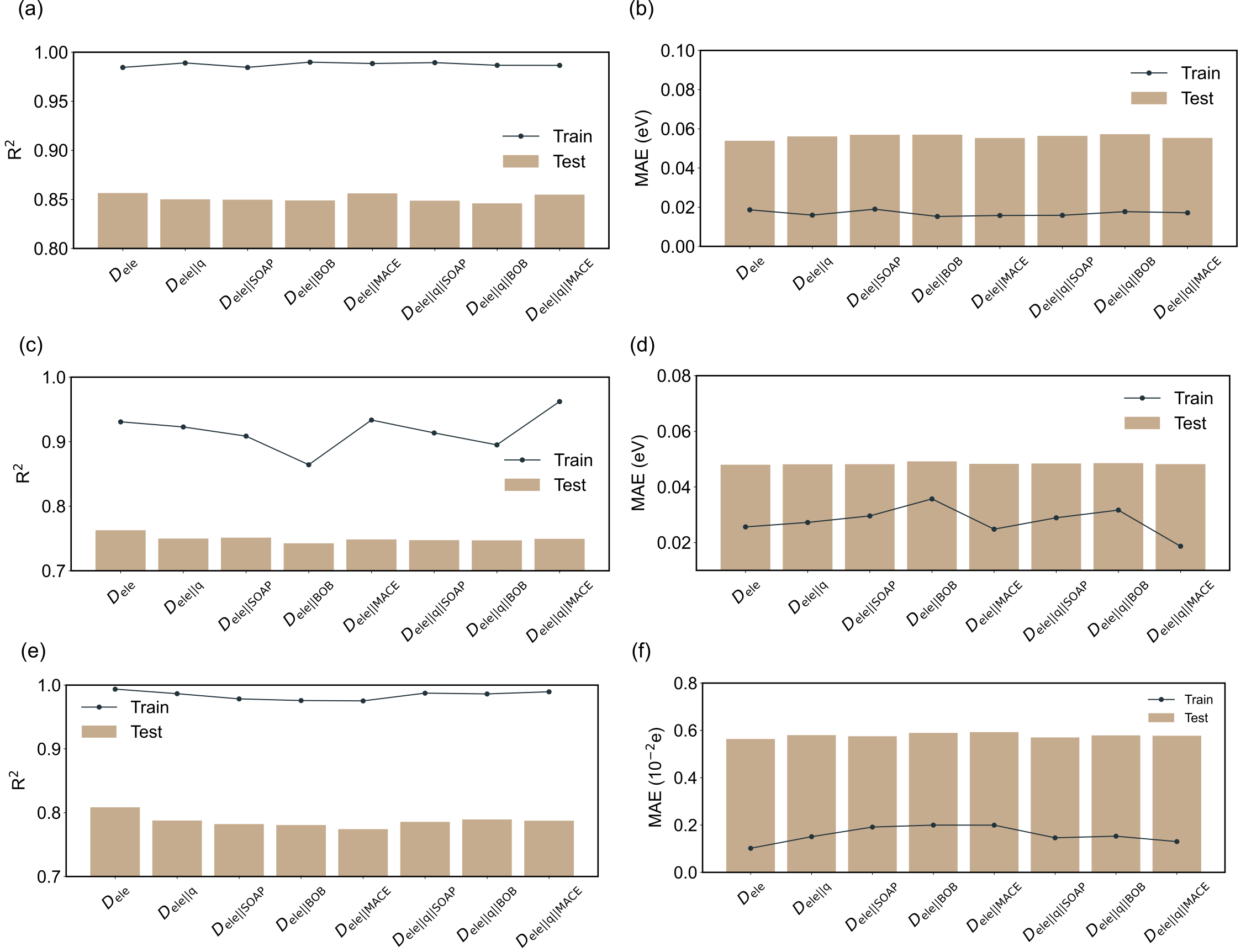}
    \caption{The model performances by combining the $D_{\rm BOB}$, $D_{\rm SOAP}$ ,$D_{\rm MACE}$, $D_{q}$ and QM properties for (a)$\sim$(b) adsorption energy $E_{\rm ads}$, (c)$\sim$(d) work function change $\Delta\phi$, and (e)$\sim$(f) charge transfer $\Delta Q$}
    \label{SI_geo_1}
\end{figure}

\begin{figure}
    \centering
    \includegraphics[width=1\linewidth]{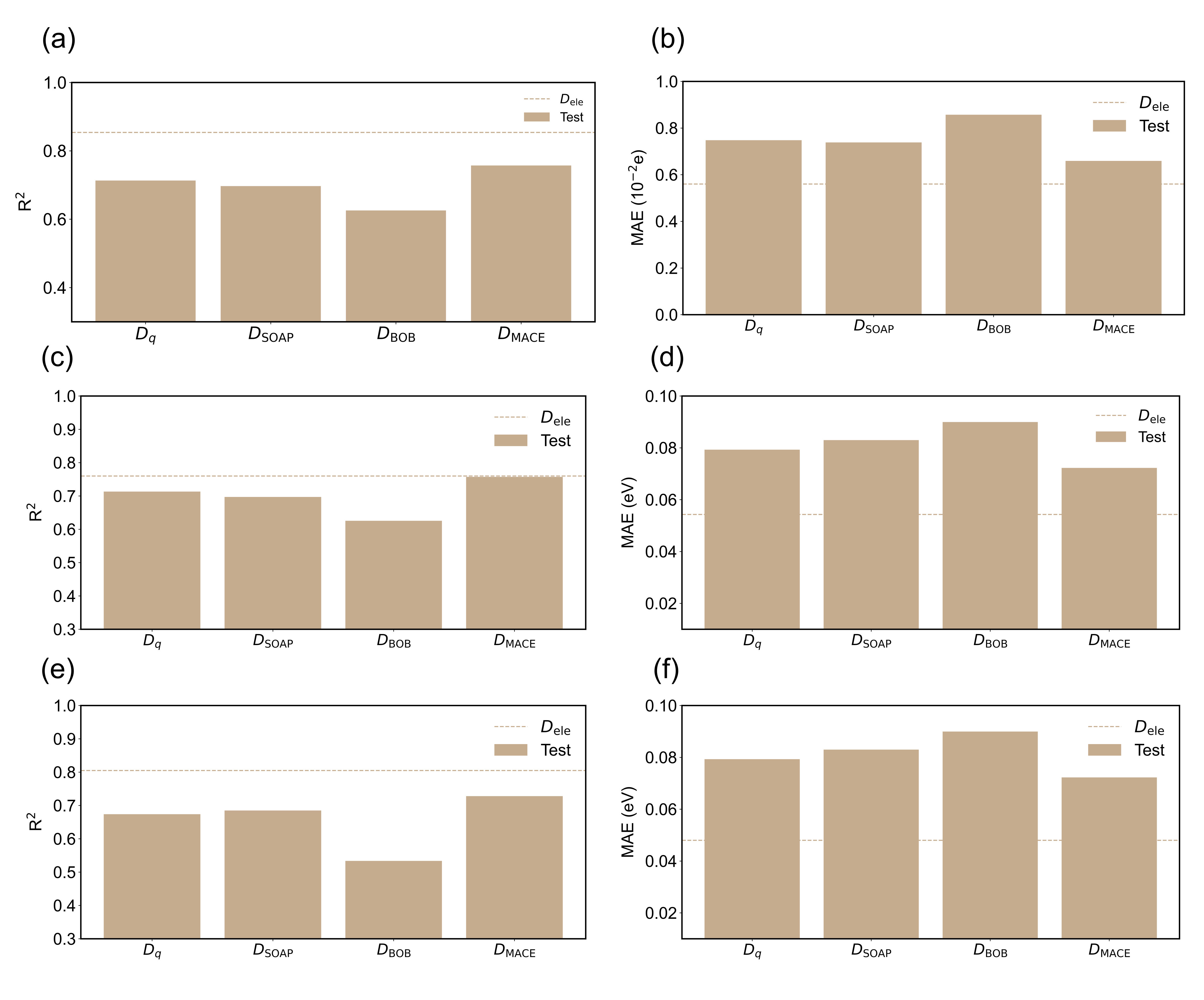}
    \caption{The same as Fig. S\ref{SI_geo_1} but only using these features without electronic properties.}
    \label{SI_geo_2}
\end{figure}

As shown in Fig. S\ref{SI_geo_1} and S\ref{SI_geo_2}, we systematically combine the trained models with the geometrical descriptors $D_{\rm geo}$ and Mulliken atomic charges $q$ to evaluate the capability of the vector features. 
In particular, the principal component analysis is applied to the geometrical descriptors to obtain the most informative expression, and accordingly, they denote ${D_{\rm BOB}}$, $D_{\rm SOAP}$ with the length of $63$ and $100$. 
And we also involved the MACE descriptor $D_{\rm MACE}$ \cite{kovacs2025mace_off} containing many-body dispersion interaction information.
In addition, the atomic Mulliken charge $D_{q}$ has been handcrafted similarly to BOB descriptors. 
And the length of Mulliken atomic charge $q$ and MACE descriptor account for $87$ and $256$, taking up of the total data number with a reasonable ratio of $\sim 4\%$.
We added these input features and retrained the models again.
The results for all binding feature predictions with $D_{\rm geo}$ and $q$ are depicted in Fig. S\ref{SI_geo_1} and S\ref{SI_geo_2}.

\clearpage
\section{Metrics for final models}

\begin{table}[htbp]
\centering
\caption{Final Catboost models metrics for binding feature prediction.}

\begin{tabular}{lccc}
\toprule
& $E_{\rm ads}$ & $\Delta \phi$ & $\Delta Q$ \\
\midrule
\multicolumn{4}{c}{\textbf{Hyperparameters}} \\
\midrule
iterations & 7000 & 5000 & 7000 \\
learning\_rate & 0.02 & 0.01 & 0.02 \\
depth & 7 & 7 & 7 \\
l2\_leaf\_reg & 9.240 & 2.387 & 1.341 \\
border\_count & 178 & 147 & 169 \\
\midrule
\multicolumn{4}{c}{\textbf{Train set}} \\
\midrule
R² & 0.984 & 0.931 & 0.994 \\
MAE & 0.019 & 0.026 & 0.001 \\
RMSE & 0.024 & 0.033 & 0.001 \\
\midrule
\multicolumn{4}{c}{\textbf{Test set}} \\
\midrule
R² & 0.857 & 0.763 & 0.808 \\
MAE & 0.054 & 0.048 & 0.006 \\
RMSE & 0.078 & 0.064 & 0.008 \\
\bottomrule
\end{tabular}
\label{tab:catboost_comparison}
\end{table}

\clearpage
\section{Explanation of work function change $\Delta \phi$ prediction}

\begin{figure}[H]
    \centering
    \includegraphics[width=0.6\linewidth]{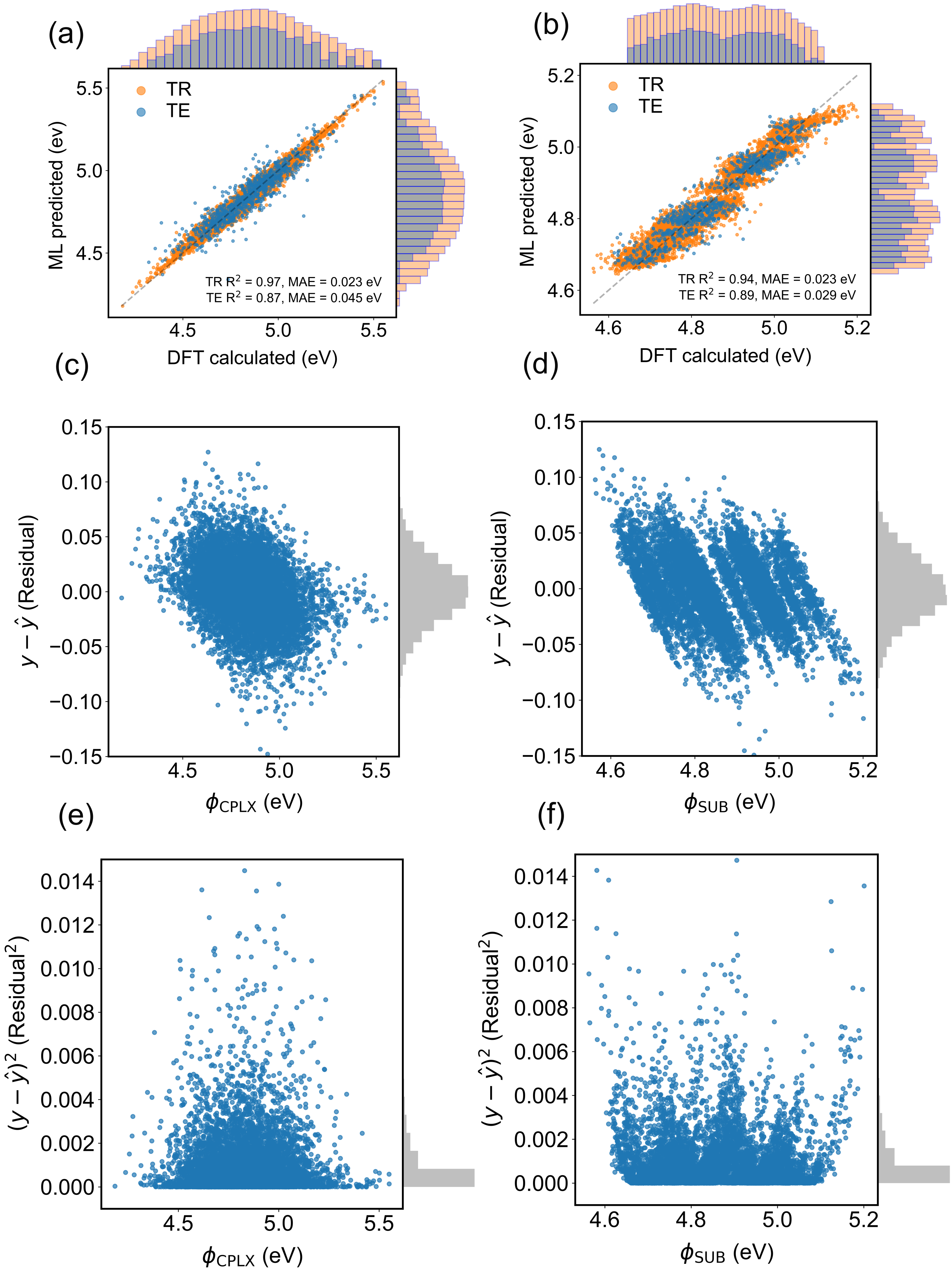}
    \caption{(a)$\sim$(b) parity plot for $\phi_{\rm CPLX}$ and $\phi_{\rm SUB}$ predictions. (c)$\sim$(d) scattering plot for residual vs. $\phi_{\rm CPLX}$ and $\phi_{\rm SUB}$. (e)$\sim$(f) scattering plot for residual's square vs. $\phi_{\rm CPLX}$ and $\phi_{\rm SUB}$.}
    \label{Work_function_predi}
\end{figure}
As shown in Fig. S\ref{Work_function_predi} (a) and (b), the individual predictions for $\phi_{\rm CPLX}$ and $\phi_{\rm SUB}$ present good performance, while the parity plot of $\phi_{\rm SUB}$ shows an unusual pattern.
To figure out for $\phi_{\rm SUB}$'s abnormal prediction behavior, we plotted the residual between the work function and the ML-predicted work function value. 
In Fig. S\ref{Work_function_predi} (c)$\sim$(d), residual vs $\phi_{\rm SUB}$ exhibit a surprisingly bunch of rod-like shapes with linear correlations, while $\phi_{\rm CPLX}$ does not show any specific pattern. 

The oscillating behavior varying from $-0.1 \sim 0.1 \rm eV$ offsets to a total tiny error when taking the residual average. Therefore, the residual square also exhibits an abnormal pattern, as shown in Fig. S\ref{Work_function_predi} (f). 
This pattern of residual of $\phi_{\rm SUB}$ might be owing to the lack of diversity of the receptor-surface leading to this systematic error, and hence imperfect prediction of the work function change $\Delta \phi$. Future work might be focused on expanding the receptor's diversity. 
\clearpage

\clearpage
\bibliographystyle{unsrtnat}
\bibliography{sample}